\documentclass[%
 aip, amsmath,amssymb,
 preprintnumbers,%
 reprint,%
]{revtex4-1}
\usepackage{lineno}
\usepackage{amssymb}
\usepackage{amsmath}
\usepackage{graphicx}

\usepackage[usenames,dvipsnames]{xcolor}%
\definecolor{goodgreen}{rgb}{0.1,0.5,0}
\definecolor{goodred}{rgb}{0.7,0,0}
\definecolor{goodblue}{rgb}{0.23,0.62,0.8}
\usepackage[colorlinks,urlcolor=goodgreen,citecolor=goodblue,linkcolor=goodred]{hyperref}
\usepackage[normalem]{ulem}

\definecolor{cream}{RGB}{222,217,201}
\usepackage{enumitem}
\newlist{todolist}{itemize}{2}
\setlist[todolist]{label=$\square$}
\usepackage{pifont}
\begin{document}

%\setpagewiselinenumbers
%\modulolinenumbers[5]
%\linenumbers
%\preprint{\dcom{Preprint not for distribution CONFIDENTIAL, \today\, Version}}
\title{Metallic Carbon Nanotube Quantum Dots with Broken Symmetries as a Platform for Tunable Terahertz Detection}
%\title{Optical properties of complex quantum dots in carbon nanotubes}

\author{G.~Buchs}
\email{g.buchs@unsw.edu.au}
\affiliation{Silicon Quantum Computing, Sydney, NSW 2052, Australia} 
\affiliation{School of Physics, UNSW Sydney, Sydney, NSW 2052, Australia} 

\author{M.~Marganska}
\email{Magdalena.Marganska@ur.de}
\affiliation{Institute of Theoretical Physics, Regensburg University, 93 053 Regensburg, Germany}

\author{J. W.~Gonz\'alez}
\affiliation{Centro de F\'isica de Materiales (CFM-MPC) Centro Mixto CSIC-UPV/EHU, E-20018 Donostia-San Sebasti\'an,  Spain}
\affiliation{Donostia International Physics Center (DIPC), Manuel de Lardizabal 4, E-20018 San Sebast\'ian, Spain}
\affiliation{Departamento de F\'isica, Universidad T\'ecnica Federico Santa Mar\'ia, Casilla Postal 110V, Valpara\'iso, Chile}

\author{K.~Eimre}
\author{C.A.~Pignedoli}
\author{D.~Passerone}
\affiliation{EMPA Swiss Federal Laboratories for Materials Testing and Research, nanotech@surfaces, \"Uberlandstra\ss e 129, CH-8600 D\"ubendorf, Switzerland}

\author{A.~Ayuela}
\affiliation{Centro de F\'isica de Materiales (CFM-MPC) Centro Mixto CSIC-UPV/EHU, E-20018 Donostia-San Sebasti\'an,  Spain}
\affiliation{Donostia International Physics Center (DIPC), Manuel de Lardizabal 4, E-20018 San Sebast\'ian, Spain}

\author{O.~Gr\"oning}
\affiliation{EMPA Swiss Federal Laboratories for Materials Testing and Research, nanotech@surfaces, \"Uberlandstra\ss e 129, CH-8600 D\"ubendorf, Switzerland}

\author{D.~Bercioux}
\email{dario.bercioux@dipc.org}
\affiliation{Donostia International Physics Center (DIPC), Manuel de Lardizabal 4, E-20018 San Sebast\'ian, Spain}
\affiliation{IKERBASQUE, Basque Foundation for Science, Euskadi Plaza, 5, 48009 Bilbao, Spain}

\begin{abstract}
Generating  and  detecting  radiation  in  the  technologically  relevant  range  of  the  so-called  terahertz  gap ($0.1 - 10$~THz) is challenging because of a lack of efficient sources and detectors.  Quantum dots in carbon nanotubes have shown great potential to build sensitive terahertz detectors usually  based  on  photon-assisted  tunnelling. A  recently  reported  mechanism combining resonant quantum dot transitions and tunnelling barriers asymmetries results in a narrow linewidth photocurrent response with a large signal-to-noise ratio under weak THz radiation. That device was sensitive to one frequency, corresponding to transitions between  equidistant  quantized  states.  In this work we show, using numerical together  with  scanning  tunnelling spectroscopy studies of a defect-induced metallic zigzag single-walled carbon nanotube quantum dot that  simultaneously  breaking  various  symmetries  in  metallic nanotube quantum dots of arbitrary chirality strongly  relaxes  the  selection  rules  in  the  electric  dipole  approximation, and removes energy degeneracies. This  leads  to  a  richer set of allowed optical transitions spanning frequencies from 1~THz to several tens of THz, for  a $\sim$10~nm  quantum  dot. Based  on  these  findings, we propose a terahertz detector device based on a metallic single-walled carbon nanotube quantum dot defined by artificial defects. Depending  on  its  length  and  contacts  transparency, the  operating  regimes range from  a  high-resolution  gate-tunable  terahertz  sensor  to  a  broadband terahertz  detector.  Our  calculations indicate that the  device  is  largely  unaffected  by  temperatures up  to  100~K, making  carbon  nanotube  quantum  dots  with  broken  symmetries  a  promising  platform  to design tunable terahertz detectors that could operate at liquid nitrogen temperatures.
\end{abstract}

\maketitle

\section{Introduction}\label{intro}
Photodetection and emission in the THz regime have recently attracted a lot of attention from fundamental and applied research communities.~\cite{Dhillon_2017} This interest is driven by the ability of THz radiation to penetrate most dielectric materials non-invasively, opening the way for numerous possible applications in the fields of medicine, security, chemical spectroscopy,~\cite{Tonouchi_2007,Ferguson_2002,Lee_2007,Mittleman_2003,Appleby_2007,Federici_2005,Siegel_2004,saeedkia2013handbook,Lewis_2019} as well as data transmission, notably in the framework of the future 6G cellular network.~\cite{Alsharif_2020} Terahertz detectors can be classified in two main categories based on their underlying detection mechanisms: thermal detectors, including bolometers,~\cite{Lewis_2019} and detectors based on collective or single electron excitation mechanisms exploiting properties inherent to quantum confinement.~\cite{Lewis_2019} 

In this context, carbon based materials such as graphene~\cite{Vicarelli_2012,Tomadin_2013,Castilla_2019,Riccardi_2020} and single-walled carbon nanotubes (SWNTs)~\cite{Wang_review:2018} have shown to be promising candidates for the next generation of THz detectors and sources. Due to their electronic properties inherited from the band structure of graphene, SWNTs fall into two classes: metallic or semiconducting.~\cite{Charlier:2007} Small bandgap semiconducting and metallic SWNTs with a curvature-induced narrow gap~\cite{Laird:2015,Portnoi_2019} are natural candidates for THz detection.~\cite{Wang_review:2018} 
Energy transitions in the THz range can be induced in metallic SWNTs quantum dots (QD) engineered with contacts~\cite{Laird:2015} or artificial defects.~\cite{Buchs:2009,Bercioux:2011,Buchs:2018,Postma_SET} Proposals for THz detectors based on QDs in SWNTs have been put forward, mostly relying on photon assisted tunneling (PAT) detection mechanisms involving excitation of electrons in the leads.~\cite{Fuse_2006,Kawano_2008,Rinzan_2012} An integrated antenna structure is usually implemented to increase the coupling efficiency between the THz radiation and the metallic SWNT QD.~\cite{Rinzan_2012} Recently, Tsurugaya \emph{et al.}~\cite{Tsurugaya_2018,Zhang_2015} performed THz spectroscopy on a SWNT QD, detecting the transition between main quantized modes of the QD through a photocurrent allowed by asymmetric tunneling barriers. They have shown that a single electron transistor (SET) geometry with nanogap metal electrodes as an integral part of a bowtie antenna can produce a narrow linewidth photocurrent response with a large signal-to-noise ratio under weak, broadband THz radiation. In their work, the metallic SWNT's linear dispersion relation gives rise to quantized states with an energy separation set by the QD length,~\cite{Buchs:2009,Ayuela_2008_a,Ayuela_2008_b} making the device sensitive to only a single THz frequency.~\footnote{In reality, the energy dispersion of a metallic SWNT marginally departs from linear,~\cite{Buchs:2009} leading to a slightly broadened set of energy transitions.} 
\\
%
%
%%%%%%%%%%%%
\begin{figure*}[!ht]
\begin{center}
\includegraphics[width=\textwidth]{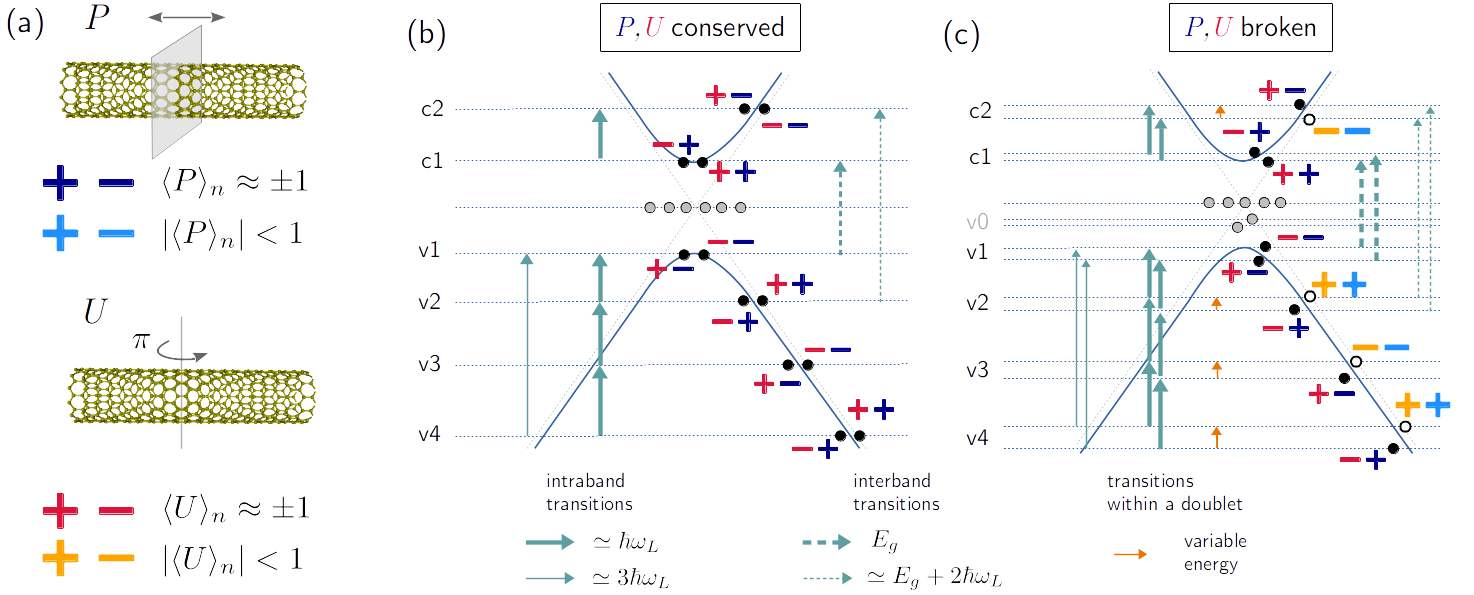}
\caption{\label{fig_one} Symmetries and optical transitions in a finite zigzag carbon nanotube. (a) $P$ denotes the reflection with respect to a plane perpendicular to the nanotube axis, $U$ the rotation by $\pi$ around an axis perpendicular to that of the SWNT. 
Each quantized state in panels (b) and (c) (labeled ``v'' for valence, ``c'' for conduction band states) has a pair of $+/-$ symbols assigned, denoting their character under both symmetry operations, with blue symbols for $\langle P\rangle_n = \langle n | P | n\rangle$ and red for $\langle U\rangle_n = \langle n| U | n \rangle$. The dark/light color corresponds to higher/lower values of $|\langle P\rangle_n|$ and $|\langle U\rangle_n|$.  (b) The curvature of the nanotube lattice opens a gap in the spectrum, thus the energy levels are not all equidistant, but most transitions nevertheless occur at close-to-integer multiples of $\omega_L = \pi v_\text{F}/L$. Edge states form within the gap, with no definite symmetry (grey markers). (c) Under the introduction of several lattice defects, $P$ and $U$ symmetries are broken, thus converting degenerate doublets into bonding and antibonding combinations of the two valleys, to which only an approximate $P$ and $U$ character can be assigned. Selection rules are relaxed and even transitions within the split doublets are allowed | c.f. Sec.~\ref{sec_optics-numerics} for further details. The eigenstate symmetries are obtained within the exact diagonalization method, see Sec.~\ref{theory} and App.~\ref{app_theo}.}
\end{center}
\end{figure*}
%%%%%%%%%%%%
%
%
In this work, we show that simultaneously breaking translational, rotational and mirror symmetries in metallic SWNT QDs leads to the  emergence of an additional energy scale in the system, which is the internal energy splitting of the hitherto degenerate valley (or isospin) doublets.
These symmetry breaking mechanisms can take place in defect-induced quantum dots. We support this with theoretical analyses based on (complementary) tight-binding and first-principle models, as well as with an experimental study via scanning tunneling microscopy and spectroscopy (STM/STS) of a $14$~nm metallic zigzag SWNT QD, defined by Ar$^+$--ion-induced defects. As a key result, we show that breaking symmetries strongly relaxes the selection rules in the electric dipole approximation, leading to a rich set of allowed optical transitions spanning frequencies from as low as 1~THz up to several tens of THz. This symmetry breaking mechanism and the set-up presented in Ref.~[\onlinecite{Tsurugaya_2018}] constitute the basis we use to propose a potential scheme for a tunable THz detection device. We show that by tuning the parameters of our proposed set-up, we can operate it either as a sensing device able to detect  weak signals at a given THz frequency, or as a broad-band THz detector operating over a set of quantum optical transitions.  Additionally, we show that the proposed device could operate well above liquid helium temperatures. 
 
The manuscript is organized as follows: in Sec.~\ref{general_optical} we give a brief overview of the electronic properties of metallic SWNTs, discussing the optical transitions between their quantized levels and how these are modified in the presence of defects. In the first part of Sec.~\ref{expvstheo}, we present low-temperature scanning tunneling spectroscopy studies of a defect-induced QD in a zigzag metallic SWNT. In the second part, we give a theoretical account of the electronic structure of the experimental QD within complementary approaches based on tight-binding and first-principle models. We explicitly evaluate the optical properties of the experimental QD in Sec.~\ref{sec_optics-numerics}. In Sec.~\ref{implementation}, we present an implementation proposal for a tunable THz detector based on the optical response of the QD introduced in Sec.~\ref{sec_optics-numerics} discussing in greater detail the various operating regimes of the proposed THz detector. A set of conclusions is given in Sec.~\ref{outlook}. We complete the manuscript with  technical appendices: in App.~\ref{app_one}, we present additional information on the experimental setup presented in Sec.~\ref{exp}. In App.~\ref{app_two}, we present in full details the results obtained with first-principle methods, whereas in App.~\ref{app_theo} we discuss the influence of defect configurations on the level splitting within the tight-binding methods. A generalization of our results to all  quasi-metallic SWNT classes is presented in App.~\ref{app_four}. Finally, in App.~\ref{AppPAT}, we present additional details on the PAT mechanism introduced in Ref.~[\onlinecite{Tsurugaya_2018,Zhang_2015}].

\section{Optical transitions in carbon nanotubes: a general introduction}\label{general_optical}

The atomic lattice of a carbon nanotube corresponds to rolled-up graphene, with the SWNT's radius and direction of rolling (so-called chirality) determining the nanotube's electronic and optical properties. The electronic spectrum of SWNTs is composed of a set of one-dimensional subbands, corresponding to the allowed values of transverse momentum.~\cite{Saito:2003} Close to the charge neutrality  point (CNP) the lowest subbands form two valleys, inherited from graphene. There, SWNTs fall into two classes, distinguished by the position of the valleys or Dirac points within the Brillouin zone and named after their most prominent members.~\cite{lunde:prb2005,marganska:prb2015,izumida:prb2015} In the zigzag class, the two valleys correspond to different transverse momenta, and both are located near the $\Gamma$ point in the SWNT's one-dimensional Brillouin zone. In the armchair class, the two valleys lie on the same transverse momentum subband, at the distance of $\sim\pm1/3$  of the Brillouin zone width from the $\Gamma$ point. The distinct implications of both classes are more evident in nanotubes of finite length.~\cite{marganska:prb2015} In the zigzag class, the two valleys are protected by the rotational symmetry, and the energy eigenstates are fourfold degenerate with regard to valley and spin degrees-of-freedom.~\footnote{The spin-orbit coupling in nanotubes results in energy splitting of $\sim 1$~meV; thus it can be neglected at the energy scales discussed here.} In the armchair class, the valleys are protected only by the translational symmetry, which is broken in a system of finite length, and each quantized longitudinal mode splits into two two-fold doublets --- symmetric and antisymmetric combinations of the two valleys, with only spin degeneracy remaining.~\footnote{The eigenstates of a finite CNT must be standing waves only in the direction parallel to the axis. They may remain travelling waves in the transverse direction. In zigzag class CNTs they are formed by the left- and right-moving states from the same valley.}  From now on we will, however, omit the spin from our discussion and consider only  the valley degree-of-freedom and degenerate or non-degenerate valley- or isospin-doublets.

In the electric dipole approximation, two main symmetries govern the optical transitions in SWNTs,~\cite{white:prb1993,damjanovic:prb1999,barros:physrep2006} as shown in Fig.~\ref{fig_one}(a): 
%
%
%%%%%%%%%%%%%%%
\begin{enumerate}
    \item[(i)] Mirror reflection with respect to a plane perpendicular to the SWNT, denoted by $P$. This symmetry is conserved only in achiral nanotubes, \emph{i.e.} armchair and zigzag;
    \item[(ii)]  Rotation by $\pi$ around an axis perpendicular to that of the nanotube, here denoted by $U$. This symmetry is conserved in all infinite nanotubes. In a clean finite lattice, it can be broken only at the edges.
\end{enumerate}
%%%%%%%%%%%%%%%
%
%
For photons polarized along the SWNT axis,~\cite{Reich_book} the optical transitions must occur between states with both different $P$ and $U$ character.~\footnote{The photons with polarization transverse to the SWNT axis can induce only inter-subband transitions, occurring at much higher energies.} The energies associated with different optical transitions reflect the features of the nanotube spectrum. In a generic metallic SWNT of length $L$, in the absence of curvature effects,~\cite{Saito:2003} the energy levels are equidistant  and all optical transitions occur at multiples of one fundamental frequency, $\omega_L = \pi v_\text{F}/L$.\\
The curvature of the SWNT lattice, which couples the $\pi$ and $\sigma$ electron bands, almost always opens a gap in the SWNT spectrum~\cite{Ando_2000,Laird:2015} and the linear dispersion turns into separate valence and conduction bands, as shown in Fig.~\ref{fig_one}(b). The quantization is not anymore equidistant in energy, with the smallest inter-level spacing occurring close to the band edges. In chiral and zigzag nanotubes, several end states form inside the gap.~\cite{ryu:prb2003,sasaki:prb2007,mananes:prb2008,izumida:prb2016} These decaying solutions have neither definite $P$ nor $U$ character.

The presence of lattice defects, such as vacancies, breaks  $P$ and $U$ symmetries as well as the rotational symmetry around the tube axis, leading to the mixing of the valleys. Additionally, defects can lead to the possible formation of QDs with broken symmetries.
As a consequence, the valley doublets split and the energy states are no longer eigenstates of $P$ and $U$. This situation is illustrated in Fig.\ref{fig_one}(c)  for a finite zigzag SWNT. Furthermore, defects can create additional states in the bandgap of the system. Broken symmetries add complexity to the electronic spectrum compared to the clean case, characterized by non-constant energy splittings in the valley doublets. This uncovers a pathway to interesting optical applications based on the possibility of detecting several THz frequencies within the same device, as we will discuss in Sec.~\ref{sec_optics-numerics}. This conceptual framework of manipulating the energy spectrum can be validated experimentally as we will show in Sec.~\ref{expvstheo} for a metallic zigzag SWNT QD. It is worth noticing that the general principles introduced so far apply to any metallic SWNT class  | achiral zigzag and armchair, as well as zigzag and armchair classes. More details are given in  App.~\ref{app_four}. 

\section{A carbon nanotube quantum dot with broken symmetries}\label{expvstheo}
In the previous section, we have argued that broken symmetries in metallic nanotube quantum dots lead to a rich electronic structure with relaxed selection rules for optical transitions. In this section, we present an experimental realization of such a QD whose energy levels structure is studied via STS, together with a theoretical analysis based on complementary tight-binding and first-principle simulations of the system.
%
%
%%%%%%%%%%%%
\begin{figure*}[!tb]
	\centering
	\includegraphics[width=0.75\textwidth]{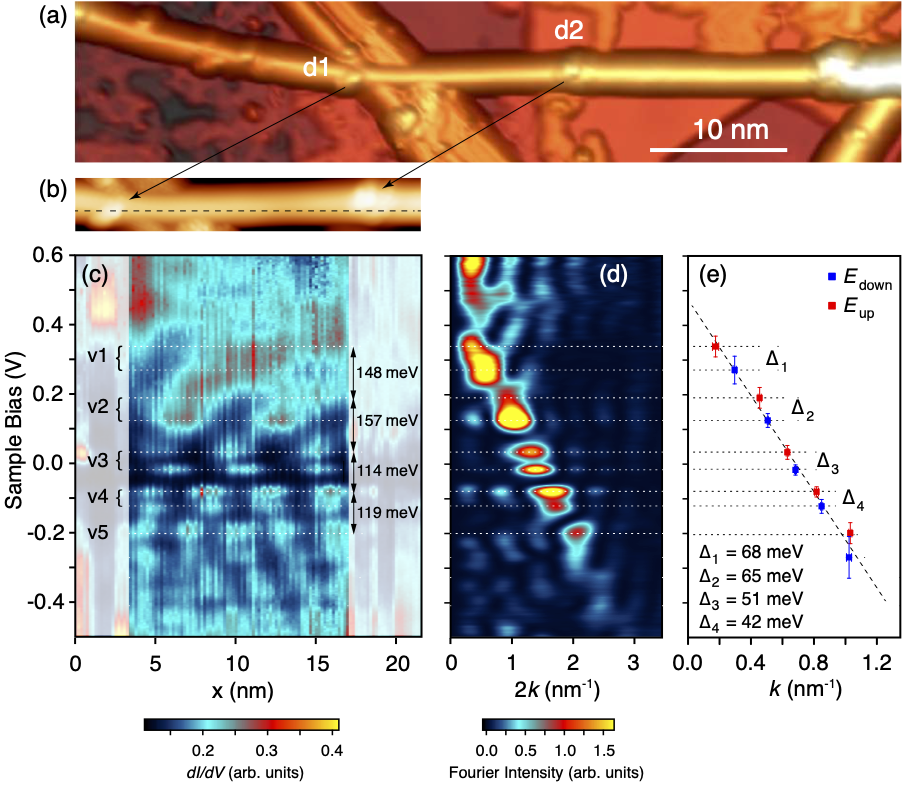} 
	\caption{\label{fig_two}(a)-(b) 3D and regular topography STM images of a quantum dot defined along a metallic zigzag carbon nanotube between Ar$^{+}$ ion-induced defects labeled $d1$ and $d2$. (c) $dI/dV(x,V)$ map recorded along the horizontal dashed line crossing defect sites d1 and d2 in panel (b). (d) $|dI/dV\left(k,V\right)|^2$ map obtained from a line-by-line Fast Fourier Transform of the $dI/dV(x,V)$ map in the interval between the transparent masks defined in the left panel. 
    (e) Energy dispersion in eV (dashed line) $E(k)=0.47\,\text{eV}-k\!\cdot\!0.693\,\text{eV\,nm}$ with the $k$ values extracted from the experimental LDOS at the energies corresponding to the split modes v1-v5  (white dashed horizontal lines) | for details c.f. App.~\ref{app_one}. STM parameters for panels (a) and (b) are: $I_{\mathrm{S}}=100$~pA, $U_{\mathrm{S}}=1$~V and respectively $I_{\mathrm{S}}=300$~pA, $U_{\mathrm{S}}=0.7$~V. STS parameters for panel (c) are: $U_\text{mod}=12$~mV, spatial resolution $x_\text{res}=1.78$~{\AA}, energy resolution~\cite{Thesis_Buchs} $\delta E\sim$ 30~meV, $T=5.3$~K. Substrate: Au(111) surface.}
\end{figure*}
%%%%%%%%%%%%
%
%
\subsection{Experimental evidence}\label{exp}
In Figs.~\ref{fig_two}(a)-\ref{fig_two}(b), we show STM topography images of a QD generated by electronic confinement between separated Ar$^{+}$ ion-induced defects (labeled d1 and d2) in a metallic zigzag SWNT. Details of the sample preparation process as well as on the STM topography and spectroscopy measurements of defects and defect-induced quantum dots are published elsewhere~\cite{Buchs:2009,Tolvanen:2009,Bercioux:2011} and summarized in App.~\ref{app_one}. The QD is embedded in a portion of the SWNT which is partially suspended between a nanotube bundle near defect site d1 and a Au(111) monoatomic terrace step (2.4~{\AA} high) near defect site d2.

In Fig.~\ref{fig_two}(c), we show a two-dimensional $dI/dV(x,V)$ map produced from consecutive $dI/dV$ spectra, proportional to the local density of states (LDOS),~\cite{Tersoff85}  recorded at equidistant locations along the dashed horizontal line in panel \ref{fig_two}(b).  The LDOS in the interval between defect sites d1-d2 reveals a series of discrete states in the voltage bias range $\sim\left[-0.4,0.5\right]$~V. Such patterns are typical of ``particle-in-a-box" states generated by confined carriers.~\cite{Buchs:2009} In contrast to the features observed in previous defect-induced QDs,~\cite{Buchs:2009,Bercioux:2011} here the quantized states are characterized by a split doublet structure with an energy separation of the order of about 40-70 meV. To gain more insight in these features, we performed line-by-line discrete Fourier transforms (see App.~\ref{app_one}) on the $dI/dV(x,V)$ map in the space interval defined between the transparent masks.~\footnote{Technical: for a better contrast of the patterns at low frequencies, the 0-frequency component (DC) is suppressed for all bias voltages. This is obtained by subtracting the average value of each $dI/dV(x,V_{j})$ line, for $V_{j}=\left[-0.5,0.6\right]$~V.} The resulting $\left[dI/dV\left(k,V\right)\right]^{2}$ map for low $k$-vectors components, corresponding to intra-valley electron scattering at defect sites,~\cite{Buchs:2009} is displayed in Fig.~\ref{fig_two}(d). It clearly shows that the third and fourth QD states (labelled v3 and v4) are split by about 51 meV and 42 meV, respectively. Further split doublet states are visible, however an accurate determination of the energy spacing is prevented by a large broadening of the states beyond the energy resolution of the setup ($\delta E \sim$ 30 meV~\cite{Thesis_Buchs}). In the following, we will refer to the higher/lower energy states of the split doublets as the up/down states. 
\\
It is worth noticing that the estimated linewidths of the states v3 and v4 up/down are about 30 and 60 meV, respectively, which is substantially smaller than the usual 100 meV or more measured with the same energy resolution (or lower) on previous metallic SWNT QDs, which were in intimate contact with the gold substrate.~\cite{Buchs:2009,Bercioux:2011} We attribute the associated longer electron lifetime to the suspended nature of the QD. This suggests that broken symmetry induced split states might also be present in previously measured metallic SWNT QDs, however their direct observation was prevented due to lifetime broadening.
\\
The energy dispersion of the SWNT QD is estimated to $E(k)=0.47\,\text{eV}-k\!\cdot\!0.693\,\text{eV\,nm}$ | Fig.~\ref{fig_two}(e). With $\hbar v_{\mathrm{F}} \simeq 0.693$~eV~nm, the Fermi velocity is evaluated to $v_{\mathrm{F}}\simeq 10.5\cdot10^{5} $m\,s$^{-1}$, in reasonable agreement with commonly reported values.~\cite{Laird:2015} With an estimated QD length of about 14.5 nm, the expected energy spacing between consecutive QD states is $\Delta E =  \hbar v_\text{F}\pi/L  \simeq 2.18/L[\text{nm}]~\text{eV} \simeq \text{150~meV}$.\cite{Buchs:2009} This is in good agreement with the experimental values measured between QD up states in Fig.~\ref{fig_two}(c). The position of the CNP is estimated to $\epsilon=0.47$~eV. This shift can finds its origin in \emph{e.g.} a charge-transfer effect due to the metallic substrate.~\cite{Clair_2011} More details on the QD structure including the determination of the SWNT chirality are presented in App.~\ref{app_one}.
\subsection{Theoretical interpretation}\label{theory}
In the following, the experimentally observed energy spectrum is investigated theoretically in detail in the context of broken symmetry relations. To this end we are using three complementary methods:
%
%
%%%%%%%%%%%%%%%
\begin{enumerate}
    \item[(i)]  The recursive Green's function (GF) technique~\cite{datta1997electronic,datta_2005}  allows us to calculate the electronic LDOS in a transport setup, with the QD seamlessly connected to semi-infinite leads. The QD is defined by the defects and/or the difference in the chemical potentials between the leads and the central suspended part. The Hamiltonian is based on a tight-binding (TB) approach.
    \item[(ii)] Exact diagonalization (ED) of the TB Hamiltonian of a SWNT segment of the same length as the experimental QD: this method gives us access to the wave functions and their character inder the $P$ and $U$ symmetries. 
    \item[(iii)] First-principles methods (FP) allow us to investigate the influence of the defects' atomic structure on the electronic spectrum of a SWNT QD with broken symmetries. In addition, this  method enables us to estimate the transparency of defect-induced tunneling barriers. 
\end{enumerate}
%%%%%%%%%%%%%%%
%
%
The GF/ED and FP methods are presented in details in App.~\ref{app_theo_1}/\ref{app_theo_2} and App.~\ref{app_two}, respectively. Here we present the insights gained within the three approaches. 
%
%
%%%%%%%%%%%%
\begin{figure*}[!t]
\begin{center}
\includegraphics[width=0.75\textwidth]{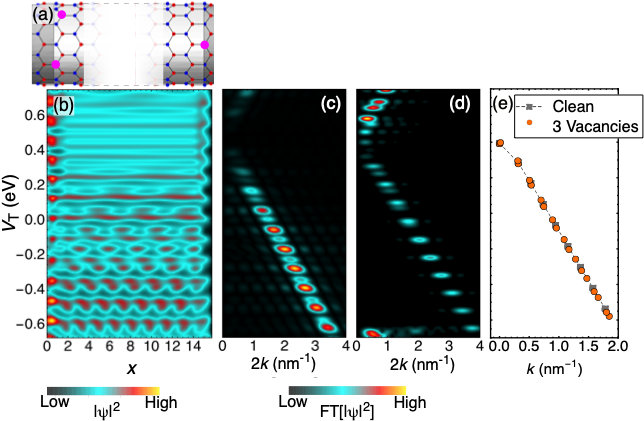}
\caption{\label{fig_three} Panel (a): Sketch of the left and right ends of a metallic zigzag SWNT QD , connected to infinite nanotube leads (shaded in grey), and defined by three impurities (magenta dots), two on the left and one on the right side of the QD. Red and blue atoms belong to the $A$ and $B$ sublattice, respectively. Panel (b): LDOS obtained in the GF approach, as a function of energy and position along the QD axis for the defect configuration in (a). Panel (c): Line by line Fourier transform of the LDOS in (b). Panel (d) Fourier transform of the LDOS obtained by first principle methods, with hydrogen adatom defects, here, we have implemented a level broadening of the FP data by $30$~meV. Panel (e):  spectrum obtained with the ED method for a finite portion of a zigzag metallic SWNT with the same length as the QD defined in (a). Energy values of a QD without (squares) and with (full circles) defects, with the corresponding wavevector on the abscissa.}
\end{center}
\end{figure*}
%%%%%%%%%%%%
%
%

The TB Hamiltonian in the single-particle approximation describing the SWNT system reads:~\cite{Charlier:2007} 
%
%
%%%%%%%%%
\begin{align}\label{H0}
\mathcal{H}_\text{SWNT}=-\sum_{\langle i,j \rangle}t_{i,j} c_i^\dag c_j + \sum_i \varepsilon_i c_i^\dag c_i,
\end{align}
%%%%%%%%%
%
%
where the sum runs $\langle \ldots \rangle$ over nearest-neighbor carbon atoms , and the operators $c_i^\dag$ and $c_i$  create or annihilate a $p_z$ electron at the lattice site $i$, respectively.  In the GF method, the hopping  parameter is constant $t_{i,j}=t\approx 2.75$~eV,~\cite{Reich2002,Charlier:2007} whereas, in the ED method the hopping  parameter depends on the atomic position and takes into account the curvature of the nanotube.~\cite{Ando_2000,delValle_2011} The second term in Eq.~\eqref{H0}  describes a modulation of the on-site energy, modelling either a local shift of the CNP (continuous modulation) or the presence of vacancy defects ($\delta$-like potential, simulating infinite barriers). In the case of a shift of the CNP we have:
%
%
%%%%%%%%%
\begin{align}\label{potential}
    \varepsilon_i=\epsilon_0 \Theta(x_i)[1-\Theta(x_i-\ell_\text{QD})]\,,
\end{align}
%%%%%%%%%
%
%
where $\ell_\text{QD}$ is the length of the QD, and $\Theta$ is the Heaviside step function. 
The value of~$\epsilon_0=0.47$~eV is set to match the change in the CNP observed experimentally | c.f. Fig.~\ref{fig_two}.
\\
In the GF approach, defects are implemented as hydrogen adatoms, described with the following Hamiltonian:
%
%
%%%%%%%%%
\begin{align}\label{imprgf}
    \mathcal{H}_\text{adatom}=\bar{\varepsilon} d_n^\dag d_n-t_\text{d}\left(c_{p_n}^\dag d_n+d_n^\dag c_{p_n}\right)\,,
\end{align}
%%%%%%%%%
%
%
where, $d_n^\dag$ and $d_n$ creates or annihilates an electron of the adatom, and $p_n$ is the position of the carbon atom closest to the adatom. In the case of hydrogen adatoms, we have used $\bar{\varepsilon}=-0.4$~eV and $t_\text{d}=5.72$~eV. In the GF approach the TB Hamiltonian describing the SWNT lattice and hydrogen adatoms is then used to calculate the LDOS of the central QD (for details see Appendix~\ref{app_theo}).\\
Within some approximation, adatoms can be considered equivalent to vacancies,~\cite{Lehtinen_2004,Krasheninnikov_2004,Gonzalez_2015,Pereira_2006,Wehling_2010} and this is how they are modelled in the ED approach, setting the on-site energy at the defects to $\varepsilon_i = 200$~eV. The full TB Hamiltonian of the isolated SWNT lattice with vacancies in the ED approach is then diagonalized numerically,\cite{Sanderson2016} providing directly the eigenstate wave functions.

%
%
%%%%%%%%%%%%
\begin{figure*}[!tb]
\begin{center}
\includegraphics[width= \textwidth]{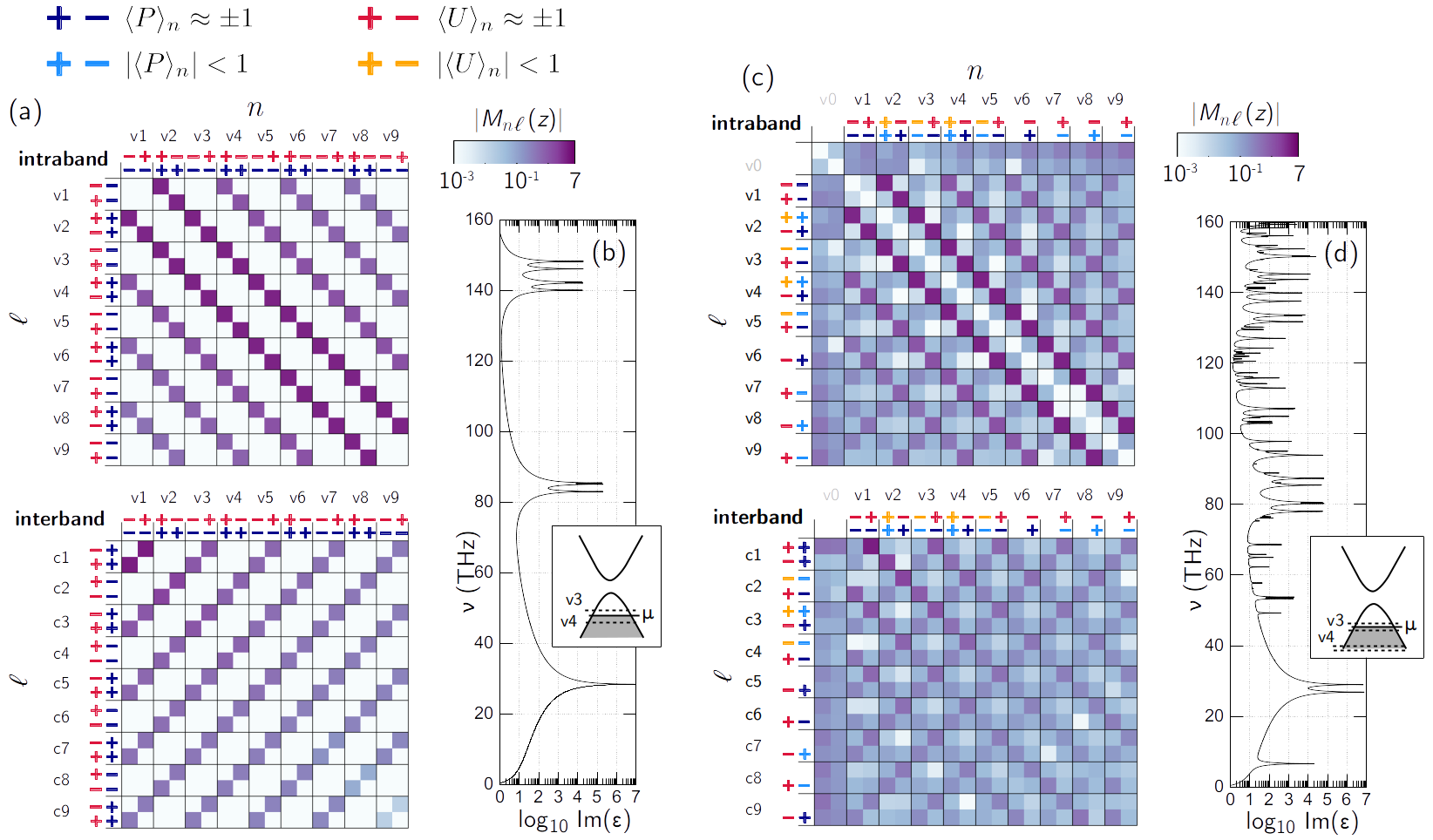}
\caption{\label{fig_transitions} Optical transition matrix elements $|M_{n\ell}(z)|$ (note the logarithmic color scale) and optical absorption curves with the chemical potential $\mu = -0.307$~eV, calculated for a (12,0) SWNT with 35 unit cells, for (a),(b) clean lattice and (c),(d) with three vacancies in the configuration described in Sec.~\ref{theory}. In both cases the length of the QD is $L$=14.9~nm. The $P$ and $U$ parity of the $n$ and $l$ states is shown as blue ($P$) and red ($U$) $+/-$ symbols. The dark symbols for symmetry $S=U,P$ denote $0.9 < |\langle S\rangle|\leq 1$, and light symbols  $0.5 \leq |\langle U\rangle|\leq 0.9$. The initial state $n$ is in the valence band. The black grid separates doublets (v$n$ for valence, c$n$ for conduction band, counted from the bandgap) of extended states. The doublet v0 forms inside the gap due to the defects and is mostly localized near them. In a clean lattice almost all transitions are forbidden, except for a few occurring between opposite $P$ and $U$ parity states. In a lattice where those symmetries are broken by the presence of three vacancies almost all transitions are allowed.}
\end{center}
\end{figure*}
%%%%%%%%%%%%
%
% 
In the experiment, the nanotube QD has a length of the order of $14.5$~nm, but for the theoretical modelling we use an integer number of zigzag unit cells leading to a theoretical QD length of $14.9$~nm corresponding to $35$ unit cells.  We can address the experimental findings of Fig.~\ref{fig_two} considering a SWNT containing three defects, two on one side of the region where we modulate the onsite energy and the other on the opposite side~\cite{Zhou:2007} | c.f. Fig.~\ref{fig_three}(a).~\footnote{In the GF method, in order to get a smooth LDOS on the surface of the carbon nanotube and account for the finite size of the STM tip, we have performed a convolution of the LDOS with a function of the form $\exp(-\lambda r)$, where $r$ is the distance of a fictitious tip of a STM from the carbon atom at $x$, and $\lambda$ is an opportune constant.~\cite{Mayrhofer:2011}} Within this configuration, we break the rotation symmetry and also the sublattice symmetry, though this can be already achieved with a single defect. In Fig.~\ref{fig_three}(b)  the LDOS evaluated within the GF approach for this configuration is displayed, and the corresponding line-by-line Fourier transform is shown in Fig.~\ref{fig_three}(c). Clearly resolved  split states with level spacing: $\Delta_3=58$~meV, $\Delta_4=51$~meV, $\Delta_{34}=114$~meV and $\Delta_{45}=115$~meV are in excellent agreement with the experimental case in [c.f. Fig.~\ref{fig_two}(d)]. In Fig.~\ref{fig_three}(d) we show the Fourier transform of the LDOS obtained within the FP method. In Fig.~\ref{fig_three}(e), we show the dispersion relation extracted from the quantized states obtained within the ED approach | for comparison, we show the case of the clean, \emph{i.e.} defect-less, nanotube as well. The results of the three methods are in good agreement with each other and with the experimental results. In the ED and FP approaches, we have included the curvature of the SWNT, and as a consequence, the spectrum presents a small deviation from the linear behavior when approaching the CNP at $\epsilon_0$. Additionally, the physical parameters in the FP and ED methods were chosen differently, leading to slightly different Fermi velocities. In spite of this, the splitting is clearly observable also within the ED and FP models.  A similar analysis of an SWNT QD without defects and with a single defect is presented in App.~\ref{app_theo_1} and~\ref{app_theo_2}  for the GF and ED methods.\\
While with other defect types the details of the spectrum would be different (cf. Fig.~\ref{fig:dft} with the results of FP calculations), the broken symmetries would remove the valley degeneracy for any defect type.

It is worth mentioning that the split states which we have experimentally and theoretically observed are substantially different in nature from the energy splittings observed in armchair nanotubes via transport spectroscopy by Sapmaz~\emph{et al.} in Ref.~[\onlinecite{Sapmaz05}]. In their case, the apparent energy splittings arise from the fact that quantized momenta are not commensurate with the position of the Dirac points, \emph{i.e.} $\pm m\Delta k = \pm2\pi m/L \neq \mathbf{\pm K}$, with $m \in \mathbb{N} $, $L$ being the length of the QD and $\pm\mathbf{K}$ being the Dirac points. Additionally, in their experiment the level spacing is constant within the linear dispersion approximation, whereas the split energies in our case originate from the breaking of fundamental symmetries in a metallic zigzag SWNT and are not constant in energy.

\section{Optical properties: numerical results} \label{sec_optics-numerics}

The optical absorption of photons with energy $\hbar\omega$ and polarization given by the vector $\mathbf{e}$, stems from the imaginary part of the dielectric function $\varepsilon(\mathbf{e},\omega)$. The optical absorption of carbon nanotubes is usually calculated under the assumption of a fully occupied valence band and an empty conduction band.~\cite{Gruneis:2003,Malic:2006} Nevertheless, the charge transfer between the SWNT and its environment (substrate and/or leads) usually results in a doping of the SWNT.~\cite{Clair_2011,Schmid:2015,Dirnaichner:2016}\\
Within the electric dipole approximation, the imaginary part of the dielectric function is obtained from the sum over all possible transitions, weighted with the Fermi's golden rule while keeping track of empty and occupied states,~\cite{Grosso:2000,Haug:2004}
%
%
%%%%%%%%%%
\begin{equation*}
\mathrm{Im}\,\varepsilon(\mathbf{e},\omega) \propto \sum_{n\ell} \frac{|\langle n |\mathbf{e}\cdot\mathbf{p}| \ell \rangle|^2}{(E_n -E_\ell)^2} \;F(\omega,E_n,E_\ell), 
\end{equation*}
%%%%%%%%%%
%
%
where in an isolated system $F(\omega,E_n,E_\ell)=[f_{\text{FD}}(E_\ell)-f_{\text{FD}}(E_n)]\,\delta[\hbar\omega - (E_n - E_\ell)]$, here $f_{\text{FD}}$ is the Fermi-Dirac distribution function. In a system coupled to the environment with  strength~$\Gamma$ the function $F$ becomes
%
%
%%%%%%
\begin{equation}
\label{eq:broadening}
 F(\omega,E_n,E_\ell) = \frac{\Gamma\,\, [f_\text{FD}(E_\ell) -f_\text{FD}(E_n)]}{[\hbar\omega - (E_n-E_\ell)]^2 + \Gamma^2}.
\end{equation}
%%%%%%
%
% 
Allowed transitions are selected by the matrix elements $\langle n |\mathbf{e}\cdot\mathbf{p}| \ell \rangle$. If the polarization vector is parallel to $x_i$ axis, these become
%
%
%%%%%%%%%%
\begin{equation}
\begin{split}
\langle n |p_i| \ell \rangle & = m\, (E_n - E_\ell) \langle n |x_i| \ell \rangle\\[2mm]
& =: m\,(E_n-E_\ell) M_{n\ell}(x_i),
\end{split}
\end{equation}
%%%%%%%%%%
%
%
where we used the commutation relation $[H,x_i] = p_i/m$, relating the matrix elements of $p_i$ to those of $x_i$. The expression for $\mathrm{Im}(\varepsilon)$ becomes
%
%%%%%%%%%%%%
\begin{equation}
\label{eq:dielectric}
\mathrm{Im}\,\varepsilon(e_i,\omega) \propto \sum_{n\ell} \;|M_{n\ell}(x_i)|^2\;F(\omega,E_n,E_\ell).
\end{equation}
%%%%%%%%%%
%
%
As mentioned above, the values of $M_{n\ell}$, and as a consequence the optical transitions in carbon nanotubes depend on the symmetries of initial and final energy eigenstates, which we obtain from the ED method. The values of $|M_{n\ell}(z)|$ for a clean (12,0) SWNT of length $L = 14.9$~nm, assuming photon polarization along the SWNT axis $z$, are plotted in Fig.~\ref{fig_transitions}(a). The colored $\pm$ symbols on the axes encode the $P$ and $U$ parity of the corresponding states. Only the transitions between states of opposite $P$ and $U$ parities are allowed, as depicted schematically in Fig.~\ref{fig_one}(b). Since in the clean nanotube the matrix elements depend only on the eigenstate symmetries, the selection rules are independent of the SWNT length.  The absorption curve calculated from Eq.~\eqref{eq:dielectric} at $\mu=-0.307$~eV (between v3 and v4 doublets) shown in Fig.~\ref{fig_transitions}(b) features three main peaks which are split several times due to non-equidistant energy levels [c.f. Fig.~\ref{fig_one}(b)]. 
\\
We will now show how the presence of defects substantially changes the situation. The symmetries fixing the selection rules in a clean lattice are now broken. This implies that almost any optical transition is allowed, albeit their matrix elements may be small, as illustrated in Fig.~\ref{fig_transitions}(c), and they now depend on the nanotube length. Importantly,  broken symmetries allow for extra intra-doublet transitions associated with frequencies of the order of a few THz | the lowest optical transition, occurring when the chemical potential is tuned into the split doublet gap, corresponds to $\sim\,8$~THz. The multiple peaks visible in the absorption curve in Fig.~\ref{fig_transitions}(d) are a manifestation of both the perturbation in the energy levels and of the relaxed selection rules. The presence of additional transitions weakens the main one, changing a single absorption peak of amplitude $10^7$ in the clean SWNT to two split peaks with an amplitude of $6\cdot 10^6$ each, while the intra-doublet transition at $\sim~8$~THz has an amplitude of $\sim10^4$. The implications of the variations in absorption amplitudes in the presence of defects are discussed quantitatively in the context of a potential  THz detector proposal in Sec.~\ref{implementation}.\\
%
%
%%%%%%%%%%%%
\begin{figure}[!tb]
\begin{center}
\includegraphics[width=0.85\columnwidth]{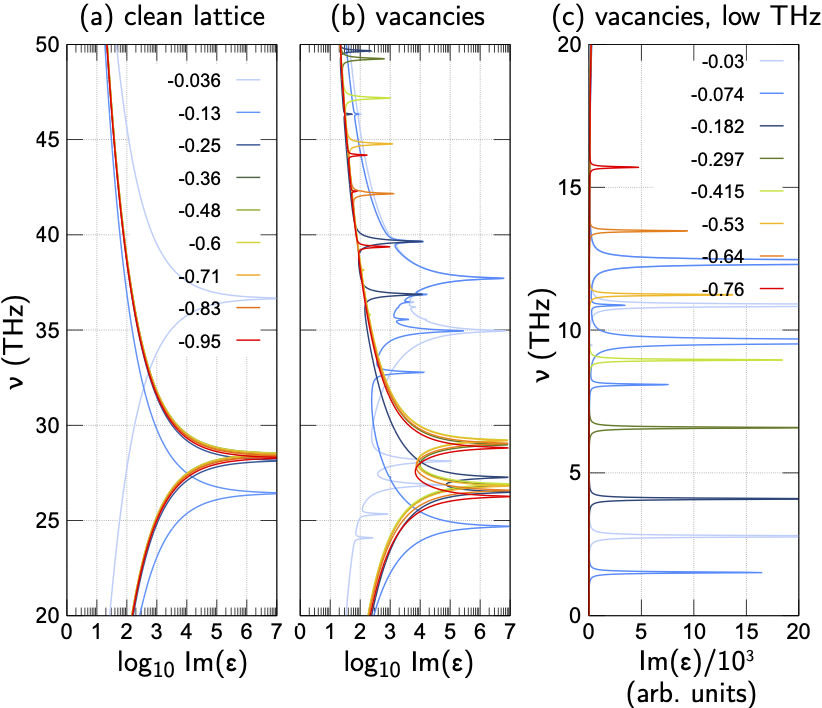}
\caption{\label{fig_tunability} Absorption curves in the THz regime for a clean (a) and defected (b),(c) (12,0) SWNT with 35 unit cells ($L=14.9$~nm), for several values of the chemical potential, given in the plots in eV units. Note that the frequency range in (c) is lower than in (a) and (b). The parameter $\Gamma$ (cf. Eq.~\eqref{eq:broadening}) is set to $10^{-4}$~eV throughout this paper.}
\end{center}
\end{figure}
%%%%%%%%%%%%
%
%
Since the breaking of symmetries does not split all doublets equally in energy, an individual SWNT QD can become sensitive to a broad set of individual THz frequencies, provided the chemical potential is tuned into the correct intra-doublet gap. 
The optical absorption curves of a (12,0) SWNT quantum dot with $L=14.9$~nm, calculated using Eq.~\eqref{eq:dielectric} from the eigenstates obtained by the ED method, are shown in Fig.~\ref{fig_tunability} for several values of the chemical potential. The values of $\mu$ are chosen to fall between the degenerate valence band doublets in a clean SWNT | Fig.~\ref{fig_tunability}(a) | and inside the split doublets in a defected SWNT | Fig.~\ref{fig_tunability}(b) | shown in Fig.~\ref{fig_one}(b),(c), respectively. The highest $\mu=-0.036$~eV in the clean case lies between the top of the valence band and the set of localized states at zero energy.
The absorption curve of a clean nanotube shown in Fig.~\ref{fig_tunability}(a) features one peak at each value of $\mu$, corresponding to the energy distance to the nearest empty level.
For the highest $\mu$ (light blue curve) the first transition (from v1 to c1) must cross the gap, hence it occurs at a higher frequency. For the next $\mu$ the first transition occurs between the v2 and v1 doublets, closest to the band gap, hence it has a much lower frequency. For decreasing $\mu$ the lowest energy transitions involve deeper v  doublets, and the absorption peaks tend to the asymptotic value of $\nu_L = \omega_L/(2\pi)$, which for $L=14.9$ nm is $\nu_{L}=v_{\text{F}} \pi/L=$~28.9~THz.  

The absorption peaks of a nanotube with the three vacancies arranged as displayed in Fig.~\ref{fig_three}(a), shown in Fig.~\ref{fig_tunability}(b), are split with respect to those in \ref{fig_tunability}(a), reflecting the removal of degeneracies. The chemical potential is now tuned to fall between the split doublet levels. The low THz part of the absorption curve in the defected SWNT, plotted in Fig.~\ref{fig_tunability}(c), shows a significant enhancement of the optical absorption at frequencies below 20 THz for several values of~$\mu$. 
The v0 doublet results from a hybridization between the SWNT and defect states, and has no symmetry. The curves for $\mu=-0.03$ (inside v0) and $\mu=-0.074$ (v1) feature several low-frequency peaks due to several allowed transitions between close doublets v0 and v1.
The first of the regularly spaced strong peaks with low frequency occurs at $\nu_1=4$~THz, $\mu_1=-0.182$~eV and the peaks for negative smaller value of $\mu$ can be observed at $\nu_n\simeq \nu_1 +n \nu_0$, with $\nu_0=2.5$~THz. Thus, by tuning the chemical potential of a single quantum dot we can access several THz frequencies. 

%%%%%%%%%%%%%%
%
%
%
%
%%%%%%%%%%%%
\begin{figure}[!tb]
\begin{center}
\includegraphics[width=1.0\columnwidth]{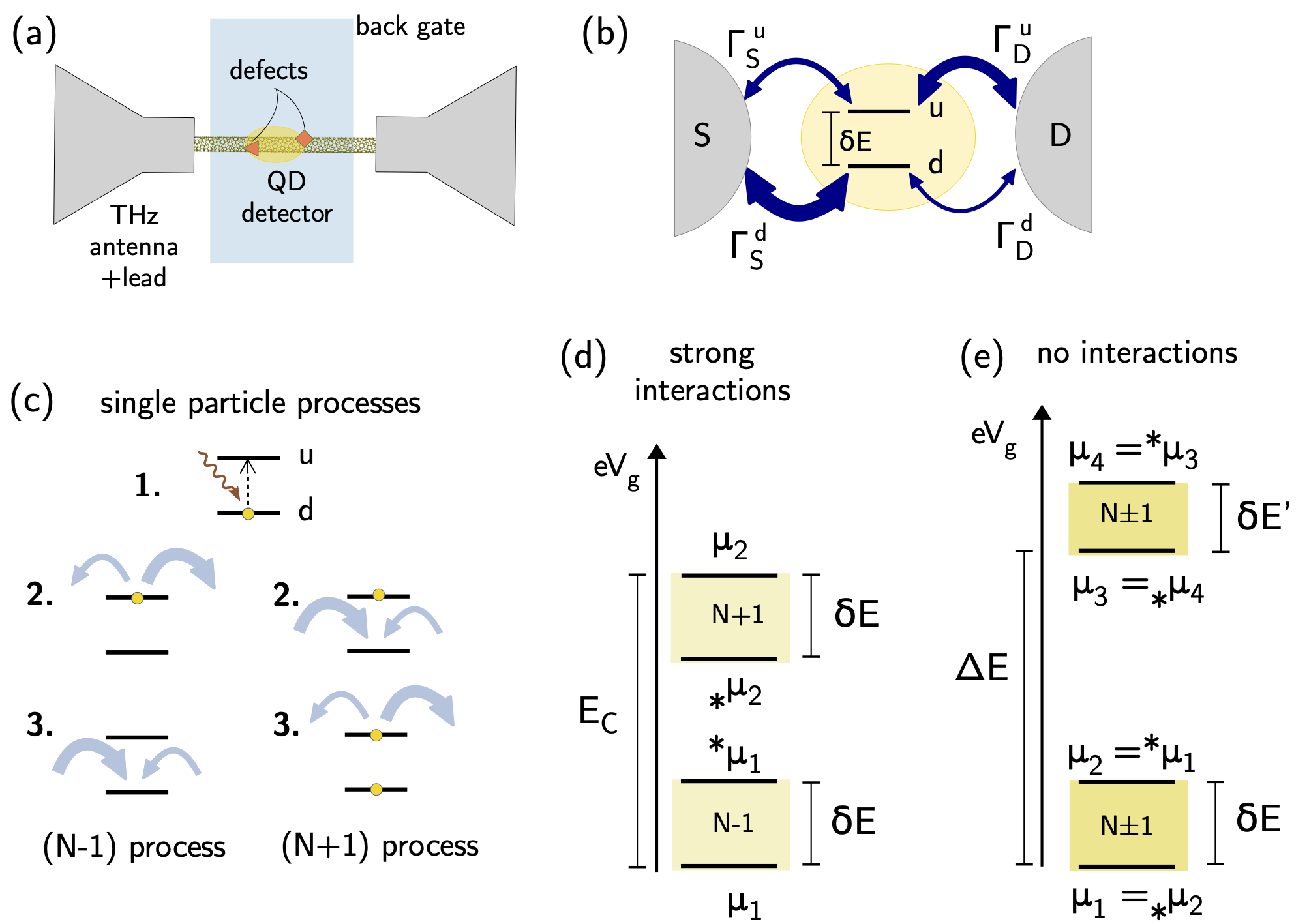}
\caption{\label{fig_implementation}(a) Sketch of the proposed device. The QD is defined by two lattice defect sites and its chemical potential is tuned with a back gate. (b) For simplicity we consider a quantum dot with one non-degenerate doublet (spin is omitted), with the $u$ and $d$ states split by $\delta E$ and coupled differently to each lead. (c) The processes leading to the appearance of the photocurrent involve a temporary emptying of the dot (N-1 process) and/or a temporarily increased occupation (N+1 process).
(d) Working ranges of the gate potential in the Coulomb blockade regime, with charging energy $E_\text{C}$ and the weak couplings obeying $\Gamma_\text{S/D}^\text{u/d} \ll \delta E < E_\text{C}$. When the gate potential is tuned so that the chemical potential of the leads falls in the shaded regions, the photocurrent can flow, either through an N-1 or N+1 process. (e) If interactions can be neglected, the dominant energy scale is $\Delta E$, the separation between doublets. In this regime $E_\text{C}=0$ and $\Gamma_\text{S/D}^\text{u/d} \ll \delta E < \Delta E$. The operational gate voltage window for the next doublet with $\delta E'$ splitting is also shown. In each window both $(N-1)$ and $(N+1)$ processes can occur.}
\end{center}
\end{figure}
%%%%%%%%%%%%
%
%

\section{SWNT Terahertz detector: proposal}\label{implementation}

Encouraged by recent formidable technological advances in high-performance SWNT devices fabrication,~\cite{Hills_2019} we propose a conceptual design for a tunable THz detector device based on a SWNT QD with broken symmetries. 
To extract a sizeable photocurrent from active transitions, we base our design on the physical mechanism described by Tsurugaya~\emph{et al.} in Ref.~[\onlinecite{Tsurugaya_2018}], allowing for high sensitivity and narrow photocurrent response. The proposed device architecture is sketched in Fig.~\ref{fig_implementation}(a) with the main element being a QD with broken symmetries defined in a metallic SWNT. The chirality can be selected using one of several recently developed procedures~\cite{SanchezValencia:2014,Yang_2015,Segawa_2016,Liu_2017,Zhang_2017,Janas_2018,Gao_2020}. In this context, it is worth emphasizing that metallic SWNTs can be obtained in suspended solutions with a degree of purity between 98~\%~\cite{MilliporeSigma_2015} and 99~\%.~\cite{nanoIntegris_2021} The SWNT containing the QD is placed between contacts incorporating a bowtie THz antenna to achieve a good coupling efficiency between the long-wavelength THz radiation and the SWNT QDs.~\cite{Tsurugaya_2018,Hirakawa_prl_2015} 
To better focus the THz light onto the antenna, a hyper-hemispherical silicon lens can be placed on the back surface of the sample.~\cite{Tsurugaya_2018,Zhang_2015} The tunability of the detector is achieved by means of a global back gate. The contacts to the external nanotube leads can be realized using state-of-the-art techniques able to suppress the Schottky barrier and lower the contact resistance below the quantum resistance. This is important to avoid embedding the QD in another, larger, QD that would be defined by the contacts.  Typical materials satisfying this condition are Mo, Pd or Au.~\cite{Nemec_2006,Cao_IBM_2015,Cao:2017} 

The working principle of our THz detector proposal shares similarities with the setup for THz spectroscopy of SWNTs in Ref.~[\onlinecite{Tsurugaya_2018}] but it provides access to a larger set of detectable frequencies and can operate in different transport regimes, determined by the coupling to the leads $\Gamma$ as  illustrated in Fig.~\ref{fig_implementation}(b)-\ref{fig_implementation}(d). For simplicity, in the following we consider only one non-degenerate doublet (spin omitted) with the higher state $u$ separated by the energy $\delta E$ from the lower state $d$. The two states couple differently to each lead | see Fig.~\ref{fig_implementation}(b) | so that a photocurrent can flow at zero bias.~\cite{Tsurugaya_2018,Zhang_2015} The functionalities of this setup depend strongly on the length $L$ of the QD and on its coupling to the leads, giving rise to trade-offs between tunability, resolution and the amplitude of the generated photocurrent. We describe the two main functionalities below. Their common features and main fabrication steps are discussed at the end of this section.
%
%
%%%%%%%%%%%%%%%%
\subsection{Short QD -- tunable THz sensor}
In order to tune the device in resonance with different THz frequencies, we need to set the chemical potential of the dot so that only the lower state $d$ is occupied and only the transition to the higher $u$ state is allowed.

%
%
%%%%%%%%%%%%
\begin{figure}
\includegraphics[width=\columnwidth]{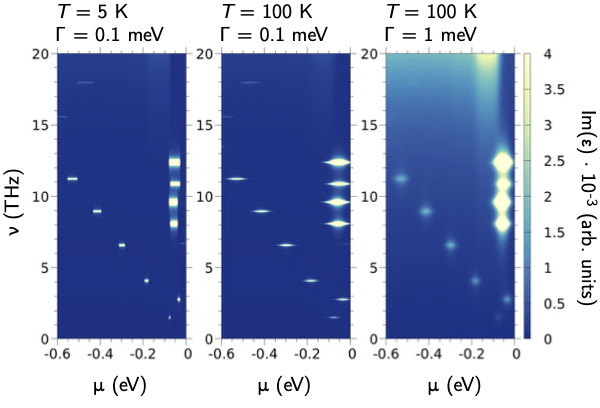}
\caption{\label{fig_continuous-mu-gate-short}
Absorption spectra in the THz regime for a (12,0) SWNT with 35 unit cells as a function of the single-particle chemical potential for some values of $T$ and $\Gamma$ (cf. Eq.~\eqref{eq:broadening}). The defect configuration is the same as in Fig.~\ref{fig_tunability}(c). For $\Gamma = 0.1$~meV, in Coulomb-dominated transport regime, the contribution of charging energy to the chemical potential has been subtracted. The color scale has been truncated at $\mathrm{Im}(\varepsilon) = 4000$~arb.~units.}
\end{figure}
%%%%%%%%%%%%
%
% 

{\em Coulomb blockade regime.|} With tunneling barriers sufficiently opaque ($R_\mathrm{t}~\gg~h/e^2$),~\cite{Tsurugaya_2018} \emph{i.e.} for
%
%
%%%%%%%%%
\[\Gamma_\text{S/D}^\text{u/d} \ll \delta E < E_\text{C},\]
%%%%%%%%%
%
%
the dot is occupied by a well defined number of electrons $N$. The rates $\Gamma_\text{S/D}^\text{u/d}$ are the tunneling rates to source (S) and drain (D) contacts for the two levels of the doublet, respectively, $\delta E$ is the splitting energy and $E_\text{C}$ is the charging energy. In this regime, the nanotube QD can be in one of four many-body states: $0,1d,1u$ or 2 electrons ($1d+1u$). The electrochemical potential for adding one electron is $\mu_1$ for $0\rightarrow 1d$ transition, $^*\mu_1$ for $0\rightarrow 1u$, $\mu_2$ for $1d\rightarrow 2$ and $_*\mu_2$ for $1u\rightarrow 2$ transition. At zero bias voltage, if the chemical potential of the dot lies inside one of the shaded ranges shown in Fig.~\ref{fig_implementation}(d), illuminating the dot with light at frequency $\nu = \delta E/h$ will promote the single electron from the $d$ to the $u$ state. From there, two processes displayed in Fig.~\ref{fig_implementation}(c) can take place.~\cite{Tsurugaya_2018,Zhang_2015} The first process called $(N-1)$ involves temporarily emptying the QD. There, the electron in the state $u$ tunnels out to the lead presenting the lowest tunnel barrier. Then the state $d$ is refilled from the leads, resetting the device. The second process called $(N+1)$ involves temporarily increasing the occupation of the QD. There, the state $d$ is filled from the leads before the electron in the state $u$ tunnels out. The final state of the QD is the same as for the $(N-1)$ process, \emph{i.e.} a single electron occupying the ground state~$d$. If the left and right tunnelling rates for either the ground or the excited state are asymmetric, the charges will flow preferentially in one direction, thus generating a measurable photocurrent.~\cite{Tsurugaya_2018,Zhang_2015} The large charging energy $E_\text{C}$ will prevent other transitions from occurring. 
Additional details of the two possible PAT mechanisms are presented in App.~\ref{AppPAT}. 

{\em Transparent contacts.|} The proposed device could also operate in a regime with transparent barriers (\emph{i.e.} $R_\mathrm{t} < h/e^2$) where the interactions are negligible | c.f. Fig.~\ref{fig_implementation}(e). In this case $E_\text{C}=0$ and the dominant energy scale is the level spacing $\Delta E$ between consecutive doublets, imposed by the size quantization,~\cite{Buchs:2009} \emph{i.e.} 
%
%
%%%%%%%
\[\Gamma_\text{S/D}^\text{u/d} \ll \delta E < \Delta E.\]
%%%%%%%
%
%
With a suppressed charging energy the chemical potentials would obey $\mu_1 =\, _*\mu_2$ and $^*\mu_1=\mu_2$, and if the gate voltage is tuned into the shaded regions, both $(N-1)$ and $(N+1)$ processes can occur. The next doublet, split by $\delta E'$, lies at an energy $\Delta E$ above, high enough to prevent its occupation. The device in this regime would again be sensitive mainly to the resonant frequency of the doublet splitting. An advantage of this setup is the large photocurrent that would be generated (higher $\Gamma_\text{S/D}^\text{u/d}$), but for the same reason the optical response peaks would be significantly broader.

\emph{Tunability.|} The absorption spectra of the short QD discussed so far, both in the Coulomb blockade regime ($\Gamma=0.1$~meV) and with transparent contacts ($\Gamma=1$~meV) are shown in Fig.~\ref{fig_continuous-mu-gate-short} for the frequency range $0\text{--}20$~THz, below the main inter-doublet transition peak at $28.9$~THz. The spectra are calculated using Eq.~\eqref{eq:broadening}, assuming that the broadening is entirely due to the contacts. By tuning the gate potential into appropriate ranges we can select the resonant frequency of the device. The tunability of the device is largely unaffected by the temperature $T$ and and coupling to the leads $\Gamma$, owing to the large energy spacing between quantized doublets in the QD.

\emph{Sensitivity.|} The magnitude of the photocurrent depends on several parameters: the antenna efficiency, coupling to the antenna leads, coupling between lead-like parts of the SWNT and the QD. We can estimate the size of the photocurrent generated by our proposed device in the Coulomb blockade regime, using as a benchmark the results of Ref.~[\onlinecite{Tsurugaya_2018}]. They recorded a photocurrent of the order of 500~pA for the main transition between two longitudinal modes in a 150~nm QD. The transition strength for a (12,0) SWNT of comparable length is of the order of $10^{9}$ in our units (cf. the caption of Fig.~\ref{fig_continuous-mu-gate-long}). The strength of the additional intra-doublet transition which we propose to exploit in our short QD is $10^{4}$ in the same units, leading to a photocurrent of the order of 5-10~fA, which is already measurable.~\cite{LTC6268-10,LTC6268,sub_pico_2017,DE-LCA-2-10} 
Increasing the coupling to the leads decreases the amplitude of the broadened absorption peaks, but enhances the current. As shown in Fig.~\ref{fig_continuous-mu-gate-short}, increasing the contact transparency does not affect tunability, therefore $\Gamma$ only drives the trade-off between resolution and amplitude of generated photocurrent.

\begin{figure}
\includegraphics[width=\columnwidth]{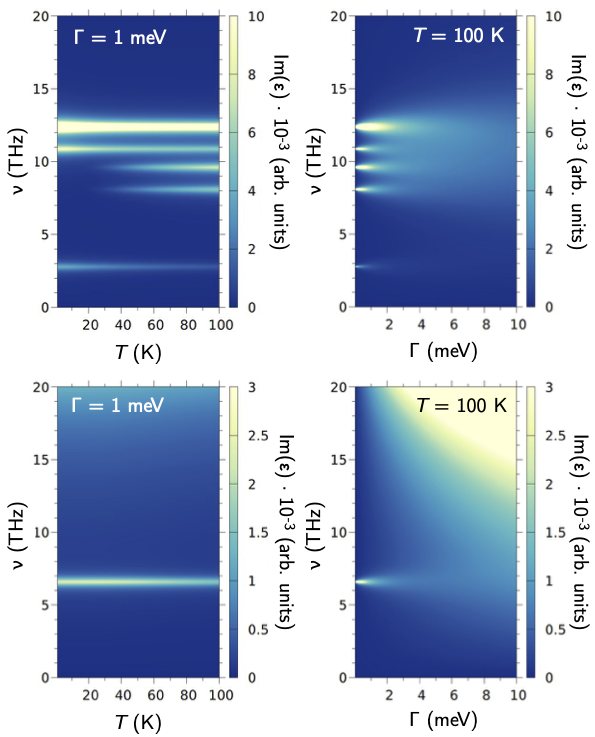}
\caption{\label{fig_T-Gamma}
Absorption spectra in the THz regime for a (12,0) SWNT with 35 unit cells as a function of $T$ (left column) and $\Gamma$ (right column, cf. Eq.~\eqref{eq:broadening}), for $\mu=-30$~meV, close to the band gap (top row) and $\mu=-297$~meV (bottom row). The defect configuration is the same as in Fig.~\ref{fig_tunability}(c), yielding several allowed transitions when $\mu$ is near the band gap, and a single sharp transition when $\mu$ lies deeper inside the valence band. For $\mu=-297$~meV the high absorption at 20 THz is caused by the broadened inter-doublet transition peak. The colour scale has been truncated for better visibility of all peaks.}
\end{figure}

\emph{Thermal range of operation.|} The linewidth of the photocurrent response peak will be determined either by the coupling to the leads or/and by the temperature. Both energy scales must not exceed our target resolution. 
Numerical absorption spectra for a (12,0) SWNT with 35 unit cells (same defects configuration as in Fig.~\ref{fig_tunability}(c)), shown in Fig.~\ref{fig_T-Gamma}, indicate that the main factor limiting the resolution is the tunnel coupling $\Gamma$, and with $\Gamma\simeq 1$~meV the absorption peaks remain sharp even at $T\simeq 100$~K. This would allow our device to operate at liquid nitrogen temperatures or with compact, commercial Stirling cryocoolers.~\cite{Clappier_1994,Air_Luquid_2021}

%
%
%%%%%%%%%%%%
%%%%%%%%%%%%
%
% 

\subsection{Open long QD -- broadband THz detector}

\begin{figure}
\includegraphics[width=\columnwidth]{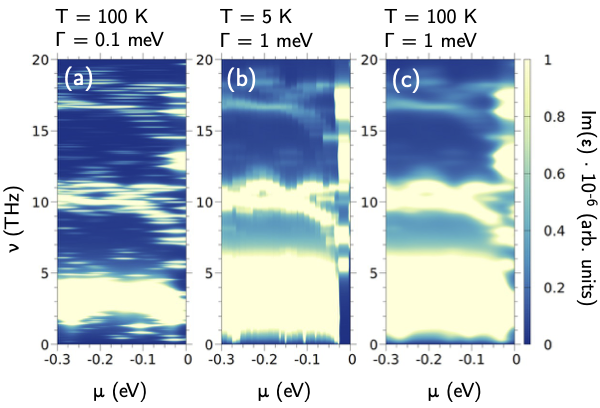}
\caption{\label{fig_continuous-mu-gate-long}
Absorption spectra in the THz regime for a (12,0) SWNT with 300 unit cells ($L=128$~nm) as a function of the single-particle chemical potential for some values of $T$ and $\Gamma$ (cf. Eq.~\eqref{eq:broadening}). The defect configuration is the same as in Fig.~\ref{fig_tunability}(c), with appropriately increased distance between left and right defects.. For $\Gamma = 0.1$~meV, in Coulomb-dominated transport regime, the contribution of charging energy to the chemical potential has been subtracted. The color scale has been truncated at $\mathrm{Im}(\varepsilon)=10^6$; the maximum absorption for the main band near $3.37$~THz is $7\cdot 10^{8}$ in (a), $1.4\cdot 10^{8}$ in (b), $1.2\cdot 10^{8}$ in (c) and (not shown) $1.4\cdot 10^9$ for $T=5$~K, $\Gamma=0.1$~meV similar to the conditions in Ref.~[\onlinecite{Tsurugaya_2018}].}
\end{figure}

When the QD is long, both the main quantization energy and the doublet splitting get smaller and the device becomes sensitive to lower frequencies. For instance, in a (12,0) SWNT with 300 unit cells ($L=128$~nm), whose absorption spectra are shown in Fig.~\ref{fig_continuous-mu-gate-long}, the frequency associated with the main quantization peak is 3.37~THz. The breaking of symmetries caused by the defects removes the degeneracies and lifts selection rules. The presence of many overlapping levels at the Fermi energy gives rise to a wide band of transitions at similar frequencies, clustering around 3.37~THz. The main band also has a weaker replica at roughly the triple of the main frequency (cf. Fig.~\ref{fig_one}). With many transitions at similar frequencies accessible in a wide range of gate voltages, the proposed device can operate as a broadband THz detector, sensitive to frequencies deep in the THz gap. In this regime we do not need resolution or tunability, and our main concern is the amplitude of the photocurrent.
With higher absorption cross-section due to greater length of the QD, also higher currents are generated. For instance, in the (12,0) QD with 300 cells, the intra-doublet transitions give rise to absorption peaks with an amplitude of the order of $10^6$ (cf. Fig.~\ref{fig_continuous-mu-gate-long}), resulting in photocurrents of the order of 1~pA. Here, both increasing $T$ and $\Gamma$ act in our favour, enhancing the range of detectable frequencies and the photocurrent amplitude.

\subsection{General discussion}
\emph{QD formation.|} The physical implementation of the QD requires a technique enabling robust asymmetric tunneling barriers separated by at least $10\text{--}15$~nm. Additionally, we require the breaking of all symmetries (translational, mirror and rotational). As discussed in Section~\ref{expvstheo}, point defects are good candidates to satisfy these conditions. These can be generated with the same Ar$^+$--ions irradiation technique used in the experimental setup presented in Sec.~\ref{exp}, calibrated for a defect separation of at least $10\text{--}15$~nm.~\cite{Buchs:2009,Buchs:2018} This technique mainly gives rise to single and double vacancies (DV) and combinations thereof.~\cite{Tolvanen:2007,Tolvanen:2009} To increase the yield of QDs with the target length and defects configurations able to break $P$ and $U$ symmetries, techniques based on focused electron beams could be used.~\cite{Robertson_2012} The resulting spectrum would depend on the type of generated defects | \emph{e.g.}, as shown in Fig.~\ref{fig:dft}, if only the conduction band could feature split doublets, this would define the operating range of our device.~\footnote{We note in passing that the defects types and their positions on the lattice could be determined from the absorption spectrum in combination with first principle simulations,~\cite{Tolvanen:2009} especially for the case of achiral metallic SWNTs (zigzag and/or armchair) with computationally tractable small unit cells.} In this context, DVs are especially interesting for a practical device due to their robustness characterized by a migration barrier of about 5 eV.~\cite{Krasheninnikov:2006} The electron scattering strength of DV's, in other words their tunnel barrier width, is energy dependent~\cite{Buchs:2018} and DV's with different orientations have different scattering profiles, leading to QDs with asymmetric tunneling barriers.~\cite{Buchs:2009,Bercioux:2011,Buchs:2018,privateLeo}  Thus, robust SWNT QDs with asymmetric tunneling barriers and all three symmetries simultaneously broken could be realized in different configurations of two DV's along the SWNT axis. This can be achieved by assuming that the angle between the DV axis (3 different orientations) and the perpendicular axis of the SWNT is different for each DV, independently of its position along the circumference of the SWNT, as shown in Fig.~\ref{fig:dft}. Simulations indicate that the resistance $R_\mathrm{t}$ of a single DV in a metallic SWNT can vary in the range $R_\mathrm{t}(\text{DV})=(0.15\text{--}0.5)~h/e^2$, depending on the energy and orientation,~\cite{Gomez-Navarro:2005,Buchs:2018} and would thus drive the QD in the transparent barriers regime (ii) described above. Note that the Coulomb blockade regime ($R_\mathrm{t} \gg~h/e^2$), has been observed in defect-induced QDs, however the exact atomic structure of the defects is unknown.~\cite{Bockrath01} Similar considerations apply also for techniques using different ion beams, such as the one based on helium.~\cite{He_2021} This method was recently employed for creating single and double vacancies in graphene,~\cite{Buchheim_2016} as well as small defects of the order of $0.5$~nm in transition metal dichalcogenides.~\cite{Mitterreiter_2020} \\
Alternatively, short QDs operating in the Coulomb blockade regime can be generated by controlled kinks produced by an AFM tip,~\cite{Postma_SET} or usual contact Schottky barriers. In the former case, we can realize in a controlled manner  QDs with a lengths of $15\text{--}20$~nm and with asymmetric coupling to left and right lead. While the kink barriers naturally break all symmetries, the Schottky barriers may need to be more carefully engineered.
%
%
%%%%%%%%%%%%
%%%%%%%%%%%%
%
% 

\emph{Influence of the substrate.|} The choice of the substrate is decisive in designing a THz detector. For our proposal, we envision three possible choices in order of complexity. The easiest solution is to use an insulating substrate, usually SiO$_2$. Devices with SWNT QDs lying on SiO$_2$ usually suffer from significant substrate- or/and process-induced disorder, especially when the electron density is reduced. However, it is worth noticing that the SWNT in the device described in Ref.~[\onlinecite{Tsurugaya_2018}] does lie on a SiO$_{2}$ substrate but shows both clear Coulomb diamonds and very narrow optical responses, suggesting that the usual SiO$_{2}$ disorder effects are still negligible at temperatures around $4\text{--}5$~K. \\
Nonetheless, in case SiO$_{2}$ related disorder would be an issue, or if better optical properties would be desirable, recent results have shown that hexagonal boron nitride (hBN) presents ideal properties as a substrate for SWNT photonic devices.~\cite{Fang_2020,Noe_2018} Notably, complex device  architectures have been fabricated with hBN as a substrate with metallic gates deposited on top.~\cite{Unuchek_2018} \\
An elegant method to avoid substrate disorder issues is to define the QD in a so-called ultra-clean suspended nanotube where all chemical processing is done before the SWNT growth.~\cite{Buchs_2011,Buchs_2014} However, this method adds an important fabrication challenge.

\emph{Plasmons.|} In the architectures described above, the  QD is definedwithin a longer portion of the metallic SWNT between the leads. The two nanotube segments that are bridging the QD and the metallic leads must be sufficiently long in order to behave like leads, \emph{i.e.} they must hold a large density of states for electrons to be able to tunnel in and out of the QD.
Let us consider here the role played by plasmons in our setup. First, the bowtie antenna enhances the THz field in the nanogap via plasmons.~\cite{Schaafsma_2013} These would be sustained by the SWNT, but because of the presence of the QD, they would mainly propagate in the two lead parts of the nanotube. If the Coulomb interaction in the leads is unscreened, and  assuming that each of these nanotube portions would have a length of the order of $500$~nm, the corresponding resonance frequency of the plasmons is estimated to be of the order of 20~THz,~\cite{Shi_2015,Morimoto_2014} \emph{i.e.} well above the ``THz gap" range which we want to address. \\

\section{Conclusions}\label{outlook}

In this work, we have shown that lattice defects can simultaneously break translational, rotational and mirror symmetries in metallic SWNT QDs leading to an irregular quantized energy spectrum characterized by states with non-equidistant energy spacings. 
We have demonstrated experimentally and theoretically that such symmetry breaking mechanisms can be achieved \emph{e.g.} in defect-induced QDs. In this context, we presented low-temperature STM/STS studies of a defects-induced QD in a metallic zigzag SWNT supported by models based on tight-binding and first-principle simulations. 

As a key result, we have shown that breaking symmetries relaxes the selection rules in the electric dipole approximation in a significant manner, leading to an extended set of allowed optical transitions spanning frequencies from about one to several tens of terahertz, therefore, well inside the so-called terahertz gap. We take advantage of these features to propose a THz detector device based on single-electron transistors similar to one presented in Ref.~[\onlinecite{Tsurugaya_2018}], thus enabling a high level of tunability and, within some parameter range, either sensing capability or a broad-band response. Importantly, our findings make carbon nanotube quantum dots with broken symmetries a promising platform to design tunable THz detectors that could operate at liquid nitrogen temperatures or using compact, commercial Stirling cryocoolers.~\cite{Clappier_1994,Air_Luquid_2021}

\section*{Author's contributions}
GB and OG conceived the experiments; GB performed the experiment and analyzed the data. AA, JWG conceived and analized data  based on the Green's function method performed by JWG. MM performed calculation based on the exact diagonalization approach, simulated and studied the optical properties of the system. KE, CAP, DP analyzed the system within first-principle methods.  GB, MM, JWG equally contributed to this work. DB coordinated the different teams efforts, provided feedback on data interpretation and coordinated the writing of the manuscript with contributions from all the authors. All the authors contributed to the preparation of the manuscript.

\section*{Acknowledgements}
The authors acknowledge useful discussions with Kazuhiko Hirakawa, Manuela Bercioux-Trummer, Geza Giedke,  Leonhard Mayrhofer, Marta Pelc, and Gabriele De Boo. M.M. thanks the Donostia International Physics Center for the hospitality. This work has been partially supported by the Spanish Ministry of Science and Innovation with PID2019-105488GB-I00 and PCI2019-103657 (A.A.) and FIS2017-82804-P (D.B.). The work of D.B. is partially supported by by the Transnational Common Laboratory \emph{QuantumChemPhys}. The Basque Government supported this work through  Project No. IT-1246-19 (A.A.).  J.W.G. acknowledges financial support from FONDECYT: Iniciaci\'on en Investigaci\'on 2019 grant N. 11190934 (Chile). A.A. acknowledge financial support by the European Commission from the NRG-STORAGE project (GA 870114). K.E., C.P. and D.P. acknowledge the Swiss National Science Foundation under Grant No. 200020\_182015 and No. 200021\_172527, and the NCCR MARVEL funded by the Swiss National Science Foundation (51NF40-182892). The Swiss National Supercomputing Centre (CSCS) under project ID s746 and s904 is acknowledged for computational resources.

\section*{AIP Publishing Data Sharing Policy}
AIP Publishing believes that all datasets underlying the conclusions of the paper should be available to readers. We encourage authors to deposit their datasets in publicly available repositories (where available and appropriate) or present them in the main manuscript. All research articles must include a data availability statement informing where the data can be found. By data we mean the minimal dataset that would be necessary to interpret, replicate and build upon the findings reported in the article. The data that support the findings of this study are available from the corresponding author upon reasonable request.

%%%%%%%%%%%%%%%%%%
%%%%%%%%%%%%%%%%%%
%%%%%%%%%%%%%%%%%%
%%%%%%%%%%%%%%%%%%
\appendix
\section{Experimental methods}\label{app_one}
\subsection{Sample fabrication}
Our sample is based on extremely pure HiPCo SWNTs,~\cite{Smalley01} with an estimated diameter distribution of 0.6-1.4 nm and a measured intrinsic defect density $<5~\mu$m$^{-1}$,  deposited on atomically flat Au(111) surfaces from a 1,2-dichloroethane suspension.~\cite{Buchs_APL_07,Tolvanen:2009,Thesis_Buchs} We irradiated the samples \emph{in situ} with $\sim$~200 eV Ar$^{+}$ ions to achieve a defect separation along the SWNTs of about 10~nm.~\cite{Buchs:2009,Tolvanen:2009}
\subsection{Sample measurement}
\begin{figure}
	\centering
	\includegraphics[width=0.75\columnwidth]{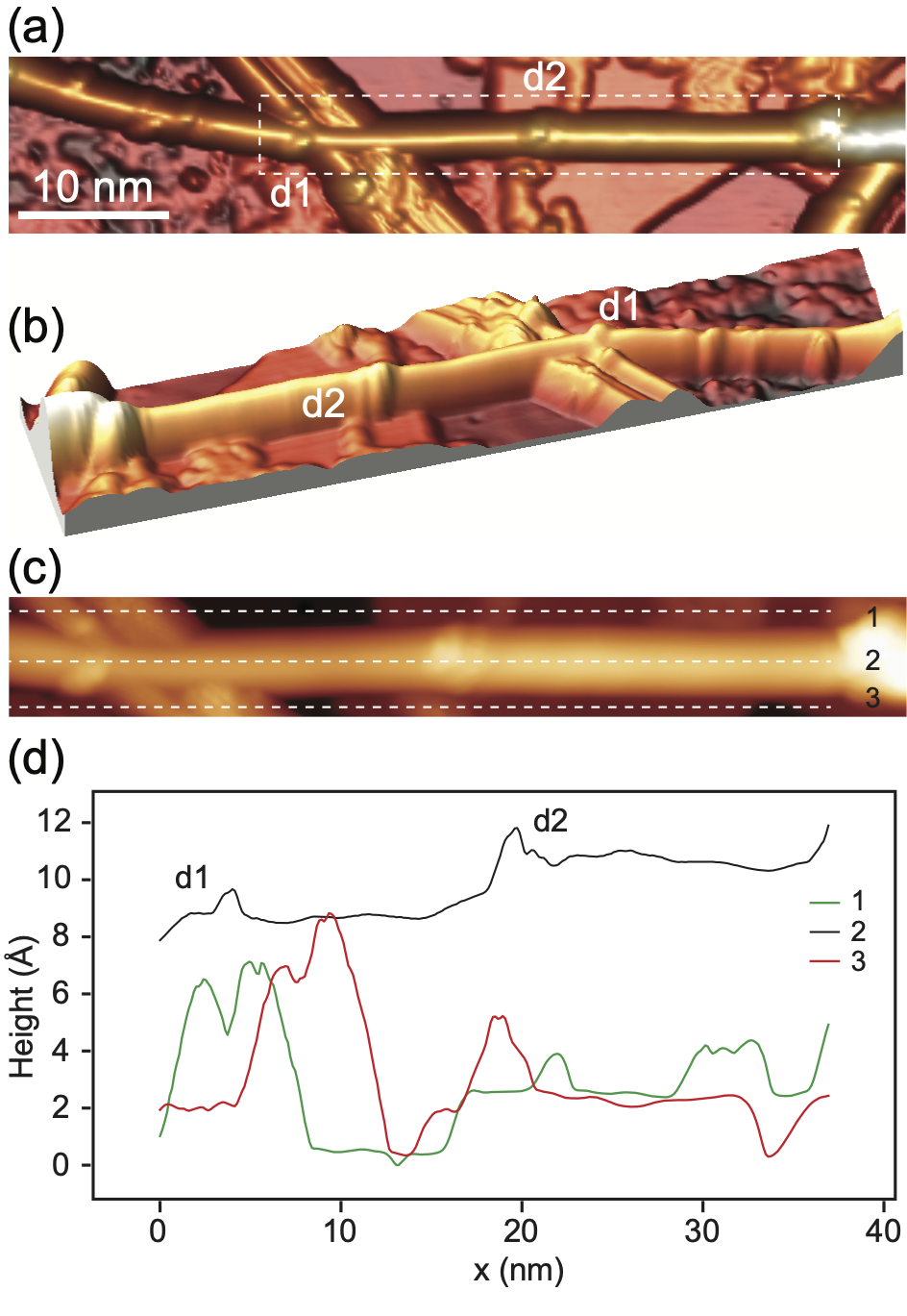} 
	\caption{\label{Sfigure:topo}(a)-(b): 3D topography STM images (processed with open source WSXM software~\cite{WSXM}) of a $\sim$ 60 nm metallic zigzag carbon nanotube portion lying on a Au(111) substrate  ($I_{\mathrm{S}}=$100 pA, $U_{\mathrm{S}}=$1 V, $T=5.3$~K). Local defects induced by collisions with $\simeq$ 200 eV Ar$^{+}$ ions appear as hillock-like protrusions. The showcased SWNT portion is crossing a nanotubes bundle in the vicinity of defect site labelled d1. In the vicinity of d2, it intersects a Au(111) monoatomic terrace step. (c): The topography image corresponds to the dashed white rectangle drawn in  (a) ($I_{\mathrm{S}}=$292 pA, $U_{\mathrm{S}}=$800 mV). (d): Height profiles along three different parallel dashed lines labelled 1 (substrate above), 2 (SWNT axis), 3 (substrate below) in  (c).}
\end{figure}

We performed scanning tunneling microscopy and spectroscopy (STM/STS) measurements  at $\sim$~5~K with a commercial (Omicron) setup operating at a base pressure $<$~10$^{-10}$~mbar. We recorded topography images in the constant current mode with  the sample grounded, using mechanically cut Pt/Ir tips. Differential conductance $dI/dV$ spectra have been recorded using a lock-in amplifier.~\cite{Thesis_Buchs,noteLDOS} All STM topography images are processed using the free software WSXM.~\cite{WSXM}
\\
Defect induced by Ar$^{+}$ ions appear as hillock-like protrusions with apparent height and lateral extension in the range 0.5-4~{\AA} and 5-30~{\AA}, respectively. In previous studies where STM/STS experiments were combined with first-principle calculations, it was shown that 200 eV Ar$^{+}$ ion irradiation mainly gives rise to single or double vacancies as well as C ad-atoms on the wall of SWNTs.~\cite{Tolvanen:2009}
\\
More detailed topography images of the SWNT configuration discussed in Sec.~\ref{exp} are shown in Fig.~\ref{Sfigure:topo}. At the crossing site, it is expected that the upper metallic zigzag nanotube portion follows the contour section of the underneath bundle with a slight compression.~\cite{Janssen02,Postma_AM00} Here, the topography image seems to indicate that the upper nanotube portion is not deformed by the bundle.
%
%
%%%%%%%%%%%
%%%%%%%%%%%%
%
%

The $dI/dV(x,V)$ maps are produced from consecutive $dI/dV$ spectra recorded at equidistant locations along a line, usually following the SWNT axis. A spatial extent mismatch between the topography image and the consecutively recorded $dI/dV(x,V)$ map of about 7\%, due to the piezoelectric voltage constant dependence on the scan velocity has been corrected by means of a compression of the $x$ scale in the $dI/dV(x,V)$ map.
%
%
%%%%%%%%%%%
\begin{figure*}
	\begin{center}
		\includegraphics[clip,width=0.9\textwidth]{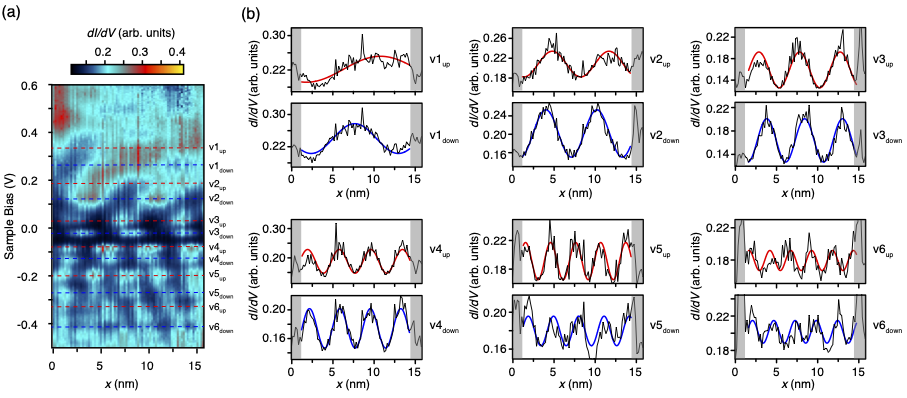}
		\caption{\label{Sfigure:linecuts} (a) $dI/dV\left(x,V\right)$ 
		map shown in Fig.~\ref{fig_two},
		(b) linecuts of $dI/dV\left(x,V\right)$ at each split doublet state (up and down).}
	\end{center}
\end{figure*}
%%%%%%%%%%%
%
%

A more detailed view on the individual quantized states is given in Fig.~\ref{Sfigure:linecuts}. Profile linecuts of the LDOS through the split doublet states are fitted with the function~\cite{Buchs:2009,Bercioux:2011} 
\begin{align}\label{fittinSin}
    |\psi(k,x)|^{2}=A+B\sin(2kx+\phi)\,.
\end{align}
The rapid modulation superimposed on the envelope of the resonant quantized states, clearly visible in panel (b), is the result of larger $k$-components corresponding to inter-valley scattering between $\mathbf{K}$ and $\mathbf{K'}$ Dirac cones whose %3-fold 
degeneracies have been lifted by the presence of the defects.~\cite{Buchs:2009,Bercioux:2011}

Line-by-line Fourier transforms are implemented with a zero padding FFT routine coded in Python: 
%
%
%%%%%%%%%%%
\begin{verbatim}
data_fft = np.fft.rfft(line,10*np.size(line,0))
norm_fft = np.abs(data_fft)**2
\end{verbatim}
%%%%%%%%%%%
%
%
with \verb?line? being a one dimensional array containing the data $dI/dV\left(x,V\right)$ for a fixed value of $V$. 
\subsection{Determination of the SWNT chirality}\label{sub3}
In practice, the $(n,m)$ chirality of SWNTs is often determined from topography measurements of chiral angles $\theta$ and STS spectra~\cite{Venema:00} compared with tables based on numerical simulations for  interaction-free SWNTs,~\emph{e.g.} Ref.~[\onlinecite{Yorikawa95}]. An unambiguous determination of the $(n,m)$ chirality is often prevented due to systematic errors of the order of $\pm$ 1$^{\circ}$ in the chiral angle measurement and renormalization of measured bandgaps due to effects inherent to the nanotube-substrate interaction.~\cite{Loiseau_renorm_2010,Buchs:2018}

\begin{figure}[!htb]
\begin{center}
\includegraphics[clip,width=0.85\columnwidth]{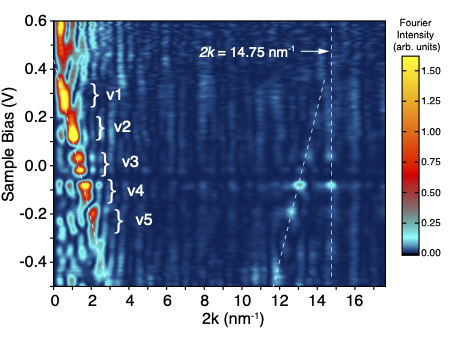}
\caption{\label{Sfigure:scattering} $|dI/dV\left(k,V\right)|^2$ map obtained from line-by-line zero padding FFT's applied on the $dI/dV\left(x,V\right)$ map in Fig.~\ref{fig_two}(b). This map is showcasing the full data set up to the Nyquist frequency $f_\text{N}=\pi/x_\text{res}$. }
\end{center}
\end{figure}

In metallic SWNT quantum dots, $[dI/dV\left(k,V\right)]^{2}$ maps, recorded inside a QD, reveal electronic scattering induced patterns where the positions of energy dependent $k$-components depend on the chiral angle $\theta$.~\cite{Buchs:2009} In other words, each chiral angle is characterized with unique patterns in the Fourier space. Figure~\ref{Sfigure:scattering} showcases the same $[dI/dV\left(k,V\right)]^{2}$ map as in Figure~\ref{fig_two}(b), but here with $k$-component values extending up to the Nyquist limit $\pi/x_\text{res}$. The presence of a single scattering cone around $2k=14.75$~nm$^{-1}$ points to a zigzag metallic SWNT~\cite{Thesis_Buchs} ($ \cos(\pi/3)|\mathbf{k}_{\mathrm{F}}| \simeq14.75$~nm$^{-1}$).
%
%
%%%%%%%%%%%%
\begin{figure}[!htb]
\begin{center}
\includegraphics[clip,width=0.75\columnwidth]{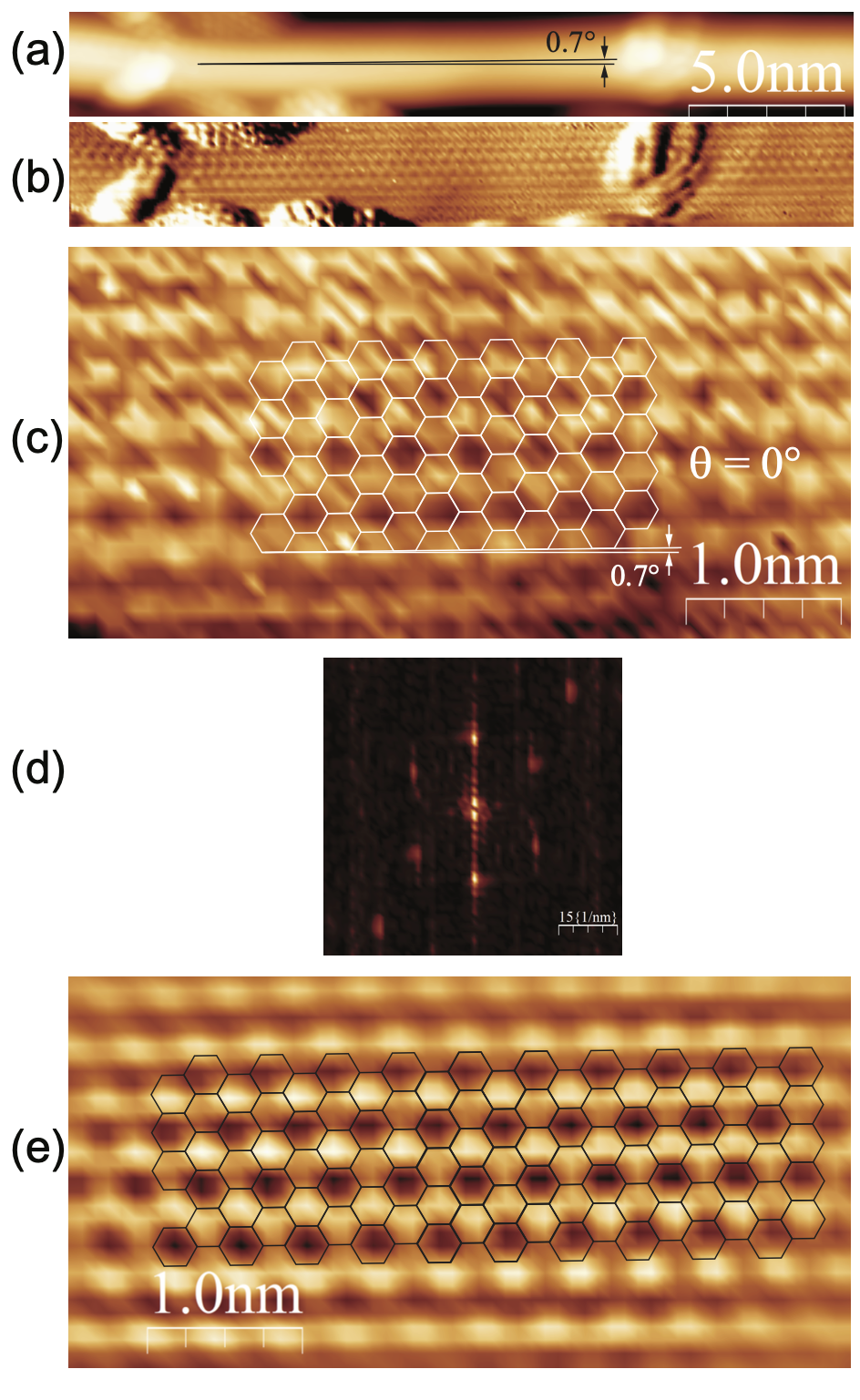}
\caption{\label{Sfigure:chirality} (a) STM topography image of the SWNT quantum dot defined by defect sites d1 and d2 ($I_{\mathrm{S}}=292$~pA, $U_{\mathrm{S}}= 800$~mV). (b) STM current image corresponding to the topography image in panel (a). (c) Zoom image taken between d1 and d2 in panel (b), with superimposed carbon-carbon honeycomb lattice. (d) 2D-FFT of the image in panel (c), with $k=2\pi/\lambda$ convention. (e) Inverse FFT image obtained from the six $k_\text{F}$ components, bandpass filtered  with identical rectangular windows (FFT tools from WSXM~\cite{WSXM}).}
\end{center}
\end{figure}
%%%%%%%%%%%
%
%
\\
Figure~\ref{Sfigure:chirality} illustrates the conventional method for chirality determination.~\cite{Venema:00} Panel (c) shows a zoomed-in region between defects d1 and d2 in the STM current image shown in panel (b), revealing a $\sqrt{3}\times \sqrt{3}R30^{\circ}$ superstructure induced by large-momentum scattering of the electrons at the defect sites d1-d2.~\cite{Buchs_APL_07} A superimposition of the graphene honeycomb lattice (in white) rotated counterclockwise by 0.7$^{\circ}$ in order to compensate for the tilt angle fits the superstructure very well. The superstructure is made even more evident after applying a narrow bandpass filter centered on the 6 Fermi vector components of the 2D FFT spectrum obtained from panel (c), followed by an inverse FFT resulting in the real space image displayed in panel (e). From these considerations, we find $\theta\simeq$~0$^{\circ}$, revealing a zigzag or close to zigzag SWNT.
%
%
%%%%%%%%%%%
%%%%%%%%%%%
%
%

From the results obtained in Figs.~\ref{Sfigure:chirality} and \ref{Sfigure:scattering}, we can confidently claim that the SWNT under investigation in Sec.~\ref{exp} is of metallic zigzag type. The diameter of the SWNT under investigation, therefore its $(n,0)$ index, cannot be reliably extracted from the STM topography image due to tip-nanotube interactions which are usually challenging to quantify.~\cite{Biro:2006} However, the central value of the estimated diameter distribution of our SWNT batch~\cite{Thesis_Buchs} reasonably points to a $(12,0)$ SWNT.
It is worth noticing that the $[dI/dV\left(k,V\right)]^{2}$ map in Fig.~\ref{Sfigure:scattering} displays a missing left-mover branch at the scattering cone compound centered around $2k=14.75$ nm$^{-1}$.  Such particle-hole symmetry breakings have been observed several times in other defect-induced metallic SWNT QDs~\cite{Buchs:2009,Bercioux:2011} and have been attributed to electron scattering selection rules linked to reconstructions between atoms from the same sublattice in double vacancy structures.~\cite{Mayrhofer:2011}
\section{First-principle investigation of complex quantum dots in carbon nanotube}\label{app_two}
%
%
%%%%%%%%%%%%
\begin{figure*}[!ht]
\begin{center}
\includegraphics[width=1.0\textwidth]{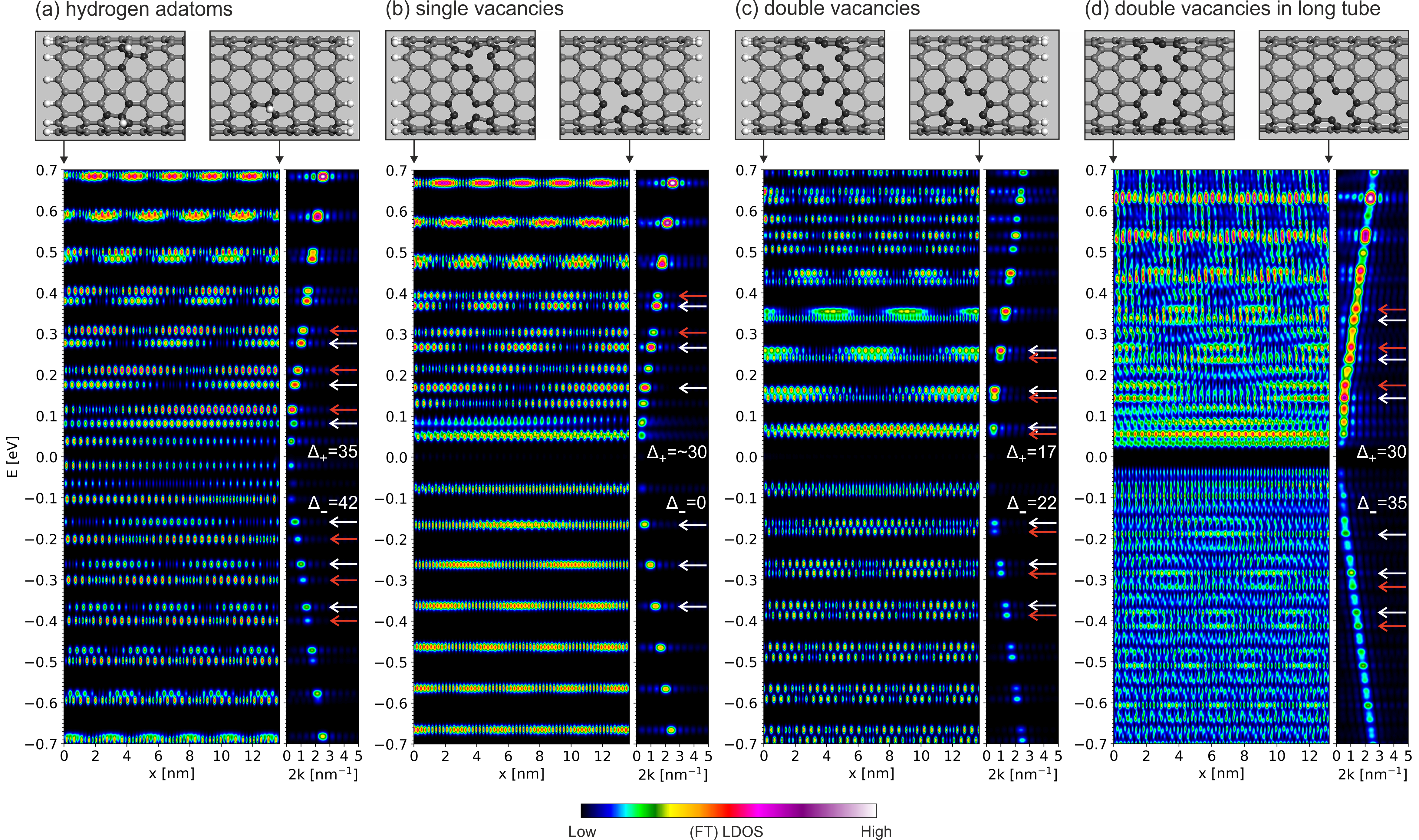}
\caption{\label{fig:dft} Density functional theory based scanning tunnelling spectroscopy calculations for carbon nanotubes with (a) hydrogen adatom defects, (b) single carbon vacancy defects and (c, d) double carbon vacancy defects. Panels (a),(b) and (c) show results for tubes which are terminated right after the defects, while in (d) the ends of the tube are $37.5$~nm away from the defects. Each panel shows the defect geometry (backside hidden for clarity; top), the LDOS between the two defected regions (bottom left) and Fourier transform of the LDOS (FT-LDOS; bottom right). White atoms are hydrogen, gray atoms are carbon and dark gray atoms show the locally relaxed carbons. In all cases, in the FT-LDOS we see bright states with a regular energy spacing of $\approx 110$~meV, corresponding to the quantized particle-in-box states between the two defected regions. Three of these states are shown by white arrows in positive and negative energies of all the systems. Due to the defects, the double degeneracy of these states is lifted and we see the partner-states at a different energy (red arrows). This energy splitting varies depending on the defect type and possibly energy position. The splitting (meV) of the closest (to CNP) identified pair of states is shown for positive ($\Delta_+$) and negative ($\Delta_-$) energies.}
\end{center}
\end{figure*}
%%%%%%%%%%%%
%
%
The electronic structure of various finite $(12,0)$ SWNTs containing different types of defects was investigated by means of density functional theory (DFT) calculations performed using the CP2K code.~\cite{Hutter2013} Specifically, we considered the following type of defects:  hydrogen adatoms, single vacancies and double vacancies. For all atomic species, we used a double zeta-valence polarization (DZVP) Gaussian basis set~\cite{VandeVondele2007} with norm-conserving Goedecker-Teter-Hutter~\cite{Goedecker1996} pseudopotentials. A 400 Ry cutoff was used for the plane wave basis set, and the Perdew-Burke-Ernzernhof (PBE) parametrization for the generalized gradient approximation of the exchange correlation functional was used.~\cite{Perdew1996} The computational cell was obtained by adding $8$~\AA\ on each side of the bounding box of the simulated system. The ends of the SWNT and the point defects show spin polarization in case of an unrestricted DFT calculation, but at the level of LDOS analysis the effect is negligible, which was confirmed for the case with hydrogen defects. Other DFT calculations were ran in the restricted Kohn-Sham formalism.

A finite clean SWNT, \emph{i.e.} without defects, was set up using an equilibrium graphene carbon-carbon distance of $1.42$~\AA\,. The carbon atoms at the end of the nanotubes were terminated with single hydrogen atoms at a distance $1.14$~\AA\,. To model the experimental system, two short SWNT portions containing the defects, separated by 35 clean unit cells of the metallic zigzag SWNT ($14.9$~nm), were introduced in the SWNT lattice. The three types of defects were positioned to match the tight-binding model shown in Fig.~\ref{fig_three}. The geometry for each defect type was first relaxed locally (up to second nearest carbon neighbors) in a short finite tube of length $2.6$~nm until forces were lower than $0.005$~eV/\AA (see geometry panels in Fig.~\ref{fig:dft} for exact defect positions and the locally relaxed atoms (dark-colored atoms)). Two different SWNT lengths were used: 
\begin{enumerate}
    \item Short SWNTs terminated by one single unit cell beyond the two SWNT portions containing the defects, to study the energy quantization and splitting patterns in a particle-in-a-box fashion;
    \item A $89.9$~nm long SWNT (211 unit cells, 10146 atoms) with the $14.9$~nm defect-induced QD in the center to study a more realistic system.
\end{enumerate}

Scanning tunneling microscopy simulations were performed using the DFT Kohn-Sham orbitals within the Tersoff-Hamann approximation, with an extrapolation of the electronic orbitals to the vacuum region to correct for the decay due to the localized basis set.~\cite{Gaspari2011,Tersoff85} The LDOS across a plane placed $4$~\AA\, above the nanotube was calculated. A weighted average with a $3$~\AA\, FWHM Gaussian filter was performed in the direction perpendicular to the tube axis to obtain a one-dimensional LDOS similar to an experimental STS line scan | Fig.~\ref{fig_two}(c). The one-dimensional LDOS was cropped to the region between the two defects and a Fourier transform was performed in order to analyze the contribution of the discrete electronic orbitals to the band structure. To remove unphysical zero-frequency modes and to increase the quality of the Fourier transform, we subtracted the average from the LDOS and zero-padded the LDOS with twelve times the length.

The LDOS and its Fourier transform for all the studied systems are shown in the various panels of Fig.~\ref{fig:dft}. The system with the hydrogen adatom defects in Fig.~\ref{fig:dft}(a) shows a degeneracy lifting of around $40$~meV for the frontier occupied and unoccupied states. This splitting gets progressively reduced and around energies $\pm 600$~meV becomes almost negligible. Same behaviour was observed for the experimental case shown in Fig.~\ref{fig_two}. The system with the single carbon vacancy defects Fig.~\ref{fig:dft}(b) represents a peculiar case, where the degeneracy lifting behaviour of occupied and unoccupied states is very different. Occupied states show no energy splitting, while unoccupied states show irregular behaviour with varying splitting (first clearly identified state-pair shows splitting of 30 meV but this is considerably reduced for the next pair). The system with double vacancy defects Fig.~\ref{fig:dft}(c) shows a very regular behaviour with an energy splitting of $~22$~meV and $~17$~meV for the occupied and unoccupied states, respectively. 

The long nanotube with double vacancy defects Fig.~\ref{fig:dft}(d) shows a much more continuous spectrum compared to the short tubes due to hybridization of the states between the defects with the other parts of the tube. However, the quasi-bound states between the two defects are still clearly visible as bright dots in the FT-LDOS, especially in the positive energies. Interestingly, the splitting is increased compared to the same defect configuration in the short tube: now the states show splitting of $~35$~meV and $~30$~meV in the negative and positive energies, respectively.
Additional results for the longer nanotube containing combinations of the considered defects show similar results. Furthermore, the optical transition matrix elements were calculated for the clean tube and the tube with hydrogen adatom defects and the resulting selection rules qualitatively confirmed the tight-binding results shown in Fig.~\ref{fig_transitions}.

\section{Tight-binding methods}\label{app_theo}
In this appendix we provide some additional technical detail related to the GF and ED methods introduced in Sec.~\ref{theory}, both based on a tight-binding approximation of the Hamiltonian describing the nanotube. Additionally, using these two methods we will analyze the cases of a SWNT with no defects or with only one defect. 
\subsection{Green's function method}\label{app_theo_1}
The Green's function method is based on the Hamiltonian describing the SWNT in Eq.~\eqref{H0}, and the term \eqref{potential} for the suspended region. Finally, the term in Eq.~\eqref{imprgf} accounts for the presence of hydrogen adatoms. These three terms describe the central part of the system $\mathcal{H}_\text{C}$. The full Hamiltonian of the SWNT QD plus contacts reads
%
%
%%%%%%%%%%%%%%%%%
\begin{equation*}
\mathcal{H}_\text{GF}= \mathcal{H}_\text{C}  + \mathcal{H}_\text{R} + \mathcal{H}_\text{L} + V_\text{L} + V_\text{R},
\end{equation*}
%%%%%%%%%%%%%%%%
%
%
where $V_\text{L}$, $V_\text{R}$ are the matrices coupling the central region with the left L and right R  semi-infinite seamless leads, described by $\mathcal{H}_\text{L/R}$. 

The Green's function of the SWNT QD is defined as:~\cite{datta1997electronic,datta_2005}
%
%
%%%%%%%%%%%%%%%
\begin{equation*}
\mathcal{G}(E) = \frac{1}{E  - \mathcal{H}_\text{C} - \Sigma_\text{L} -\Sigma_\text{R}},
\end{equation*}
%%%%%%%%%%%%%%
%
%
where $\Sigma_\text{L/R}$ are the self-energy of the left and right lead, respectively. The self-energies are evaluated as $\Sigma_\ell= V_{\ell \text{C}}\; g_\ell \;  V_{\ell \text{C}}^\dagger$ with $\ell=\text{L,R}$, and $g_\ell = (E -\mathcal{H}_\ell)^{-1}$
is the renormalized Green's function of the semi-infinite lead $\ell$. 

In order to compare with the experimental results of the STM, we evaluate the LDOS via the retarded  Green's function matrix element,~\cite{economou1984green} in a particular position of a carbon atom $x$:
%
%
%%%%%%%%%%%%%%
\begin{equation*}
\mathcal{D}(x,E)=-\frac{1}{\pi}\operatorname{Im}\left[\langle x|\mathcal{G}(E)|x\rangle\right],
\end{equation*}
%%%%%%%%%%%%%
%
%
where $\text{Im}[...]$ denotes the imaginary part. In order to get a smooth LDOS on the surface of the carbon nanotube and account for the finite size of the STM tip, we have performed a convolution of the LDOS trace along the SWNT with a function of the form $\exp(-\lambda r)$, where $r$ is the distance of a fictitious tip of a STM from the carbon atom at $x$, and $\lambda$ is an opportune constant.~\cite{Mayrhofer:2011}
%
%
%%%%%%%%%%%%
\begin{figure*}[!t]
\begin{center}
\includegraphics[width=0.95\textwidth]{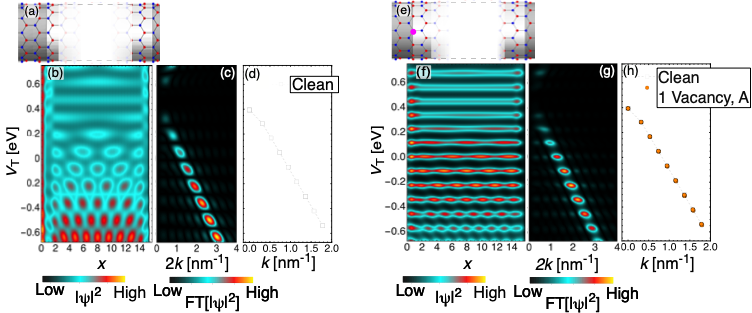}
\caption{\label{fig:two} (a): Sketch of the left and right end of a metallic zigzag SWNT with a suspended section and without impurities. (b) LDOS as a function of the position along the tube and of the energy for the case in (a). (c) Fourier transform of the LDOS in (b). (d) Wave function spectrum for a finite portion of tube with the same length as the suspended region in (a). (e): Sketch of the left and right end of a metallic zigzag tube with a suspended section and a single defect of the left side. (f) LDOS as a function of the position along the tube and of the energy for the case in (e). (g) Fourier transform of the LDOS in (f).  Panel (h) Wave function spectrum for a finite portion of tube with the same length as the suspended region in (e). The splitting of the levels is barely visible in (h) and not resolved in the FT of the LDOS in Panel (g).}
\end{center}
\end{figure*}
%%%%%%%%%%%%
%
%
\subsection{Exact diagonalization method}\label{app_theo_2}
The matrix elements of the tight-binding Hamiltonian Eq.~\eqref{H0} are calculated using the expression for the hopping between nearest-neighbour atoms $i$ and $j$ in Ref.~[\onlinecite{delValle_2011}],
%
%
%%%%%%%%%%%%%%%%%%%
\begin{equation*}
t_{i,j} = V_\pi\cos(\varphi_i-\varphi_j) - (V_\sigma-V_\pi) \frac{R^2}{a_\text{C}^2} \left[1- \cos(\varphi_i-\varphi_j)\right]^2,    
\end{equation*}
%%%%%%%%%%%%%%%%%%%
%
%
with the hopping integrals $V_\sigma = 6.38$~eV and $V_\pi = -2.66$~eV,\cite{Tomanek1988} $R$ the nanotube radius, $a_\text{C}=1.42$~\AA\, the carbon-carbon nearest neighbour distance and $\varphi_i$ the polar coordinate of the atom $i$. Without defects the on-site energies $\varepsilon_i$ are zero, in the case of lattice with vacancies $\varepsilon_i = 200$~eV at the vacancy sites and zero otherwise.\\   
The numerical diagonalization of the Hamiltonian matrix yields directly its  eigenvectors, $\boldsymbol{\psi}_{n} = \{ \psi_{n,i}\}$. The eigenstate
$|\psi_n\rangle$ of the finite lattice is constructed as $|\psi_n\rangle = \sum_i \psi_{n,i} |\mathbf{R}_i\rangle$, where $i$ runs over all atoms in the finite SWNT lattice of the central QD and $|\mathbf{R}_i\rangle$ represents the $p_z$ orbital at the site $\mathbf{R}_i$. The action of the reflection symmetry $P$ and perpendicular rotation $U$ is to permutate the set of $\mathbf{R}_i$, mapping $\mathbf{R}_i \rightarrow\mathbf{R}_{Pi}$ and $\mathbf{R}_i\rightarrow\mathbf{R}_{Ui}$, respectively. The symmetries of the eigenstates are then calculated from the scalar products
%
%
%%%%%%%%%%%%%%%%%%%%
\begin{equation*}
\langle \psi_n | S | \psi_n \rangle = \sum_{i} \psi_{n,i}^* \psi_{n,Si},    
\end{equation*}
%%%%%%%%%%%%%%%%%%%%
%
%
where $S=P,U$. The matrix elements $M_{n\ell}(z)$, determining the optical transitions for photons polarized along the SWNT axis $z$, are calculated as
%
%
%%%%%%%%%%%%%%%%%%%
\begin{equation*}
M_{n\ell}(z) = \langle \psi_n | z | \psi_\ell \rangle = \sum_i z_i \psi_{n,i}^* \psi_{\ell,i},   
\end{equation*}
%%%%%%%%%%%%%%%%%%%
%
%
with the origin of the $z$ axis set in the geometric centre of the nanotube lattice.

\subsection{Additional results within the Green's function and exact diagonalization  approaches}\label{app_theo_3}

We show here the case of a finite suspended portion of SWNT and check that for this specific case the degeneracy of the metallic mode is preserved. In the Fig.~\ref{fig:two}(a) to \ref{fig:two}(d) we present the case where the onsite energy is spatially modulated in order to model the suspended part of the SWNT and no structural defects are introduced | clean case. Specifically, we have set an onsite energy producing a change the in CNP compatible with the experimental observation of $\epsilon_0$. This specific onsite energy configuration, depicted in the Fig.~\ref{fig:two}(a) as a change of color between the two contacts, does not break rotation symmetry, thus it does not lift the twofold degeneracy of the metallic mode. This is quite clear in the LDOS and in its Fourier transform obtained via GF method, shown in Figs.~\ref{fig:two}(b) and \ref{fig:two}(c), respectively. This result is confirmed in the ED spectrum of a finite portion of SWNT corresponding to the suspended part [Fig.~\ref{fig:two}(d)]. Comparing Figs.~\ref{fig:two}(b) and \ref{fig:two}(d), we see clearly the role of contacts in broadening the system's energy levels.

In the Figs.~\ref{fig:two}(f) to \ref{fig:two}(g) we consider the presence of a single defect on the left side of the suspended region placed on an A carbon atom of the nanotube lattice. In this case we expect a breaking of the rotation symmetry and a lifting of the twofold degeneracy of the metallic mode. This can be observed in the Fig.~\ref{fig:two}(h), where the tiny splitting of the discrete energy levels obtained by the  wave function method can still be distinguished. However, this splitting is completely hidden by the presence of the contacts in the case of the Green's function method shown in Fig.~\ref{fig:two}(b). The contacts are inducing a level broadening that is larger that the splitting energy due to the lifting of the symmetry.\\
Note that with such short lattice fragments the level splitting is highly sensitive to the placement of the single defect. The left edge is formed by only $B$ sublattice atoms and the $A$ component of the unperturbed wave function has very low amplitude near the left end of the SWNT.
A defect on a $B$ sublattice at the left end (or on $A$ sublattice on the right end) would produce a larger splitting, though still smaller than that observed in the experiment. 
\section{Optical absorption in nanotubes of non-zigzag chiralities} \label{app_four}
\begin{figure}[!tb]
\begin{center}
\includegraphics[width=0.85\columnwidth]{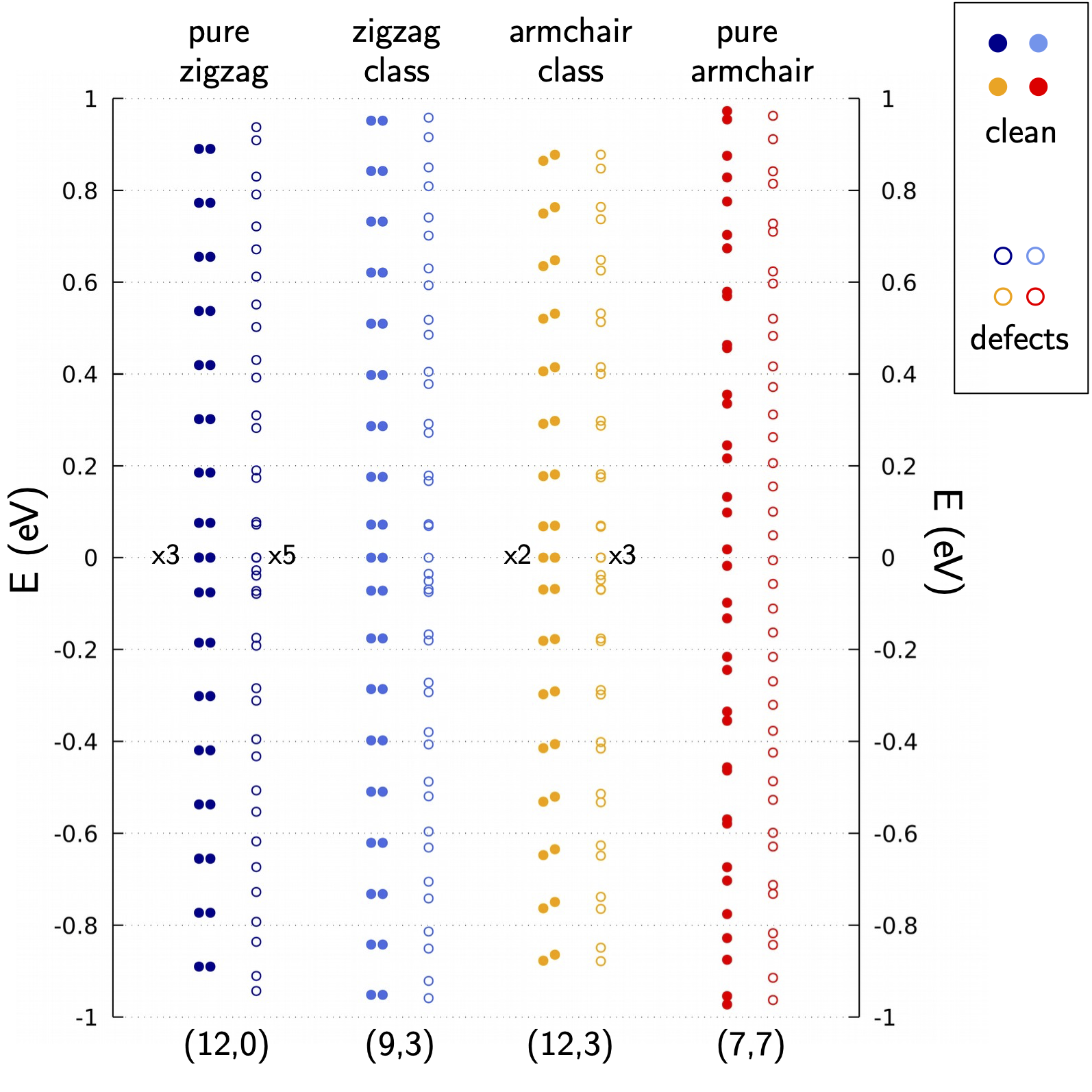}
\caption{\label{fig_spectra-chiralities} Energy spectra of isolated finite SWNTs in four different chirality classes, with approximately the same length of $\sim15$~nm. Bullets mark the energy levels of a clean lattice, open circles the levels of a lattice with three vacancies: two near the left end on A and B sublattice, one near the right end on the B sublattice. The number of doublets (for clean lattice) or states (for lattice with vacancies) at zero energy is indicated next to the appropriate marker. }
\end{center}
\end{figure}

\begin{figure*}[!tb]
\begin{center}
\includegraphics[width=\textwidth]{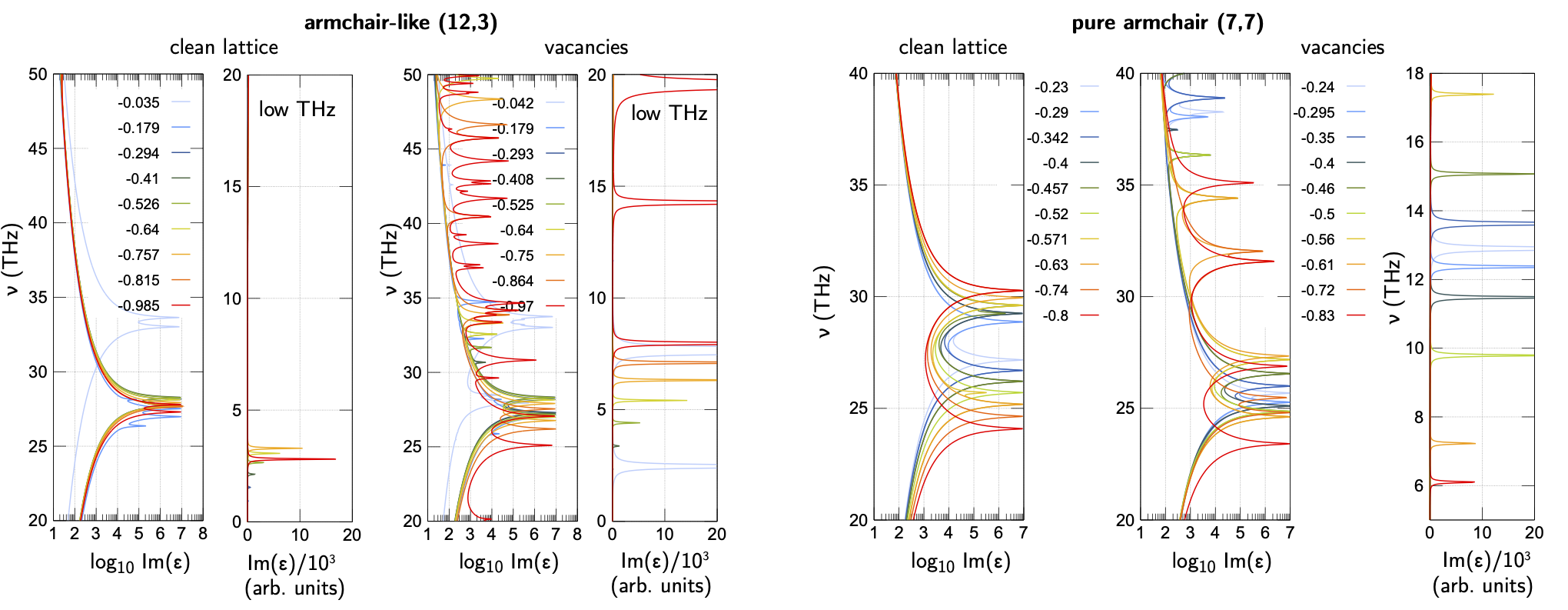}
\caption{\label{fig_absorption-armchairs} Absorption curves in the THz regime for an armchair-like (12,3) SWNT with 23 unit cells,
and for a pure armchair (7,7) SWNT with 61 unit cells, for several values of the chemical potential, given in the plots in eV units. The lack of reflection symmetry in chiral tubes allows the intra-doublet transitions to occur even in the clean (12,3) SWNT, manifesting as weak absorption peaks in low THz range. The parameter $\Gamma$ (cf. Eq.~\eqref{eq:broadening}) is set to $10^{-4}$~eV.}
\end{center}
\end{figure*}

As we have stated in Sec.~\ref{general_optical}, the optical response of a QD created on an SWNT of any chirality follows from the same physical mechanisms. The two requirements for the manifestation of low-THz absorption peaks are (i) lifting of the level degeneracy and (ii) breaking of symmetries, which allows the optical transitions between the doublet states to occur. In the Fig.~\ref{fig_spectra-chiralities} we show the energy spectra (obtained numerically) of four isolated SWNTs of different chiralities, with similar lengths of $\sim$15nm and minimal boundaries. Together they represent all chirality classes.~\cite{marganska:prb2015} The clean lattice spectra of pure zigzag and zigzag class SWNTs at low energies are doubly degenerate. In these chirality classes the defects provide both a removal of doublet degeneracy (by breaking the rotational symmetry) and a breaking of the $U$ (and $P$, in pure zigzag) symmetry. In the armchair-related classes, where the valleys are protected not by rotational but by translational symmetry, the finite nature of the lattice is enough to mix the valleys and lift the level degeneracy. The defects are however still necessary, since by breaking the $U$ (and $P$, for pure armchair) symmetries they allow intra-doublet transitions to occur. \\
The optical absorption of an armchair-class and pure armchair in Fig.~\ref{fig_absorption-armchairs} for short SWNTs shows very similar features to that of the pure zigzag presented in the main text. The lifting of the valley degeneracy in a clean lattice is visible through the presence of double absorption peaks for the clean lattices in Fig.~\ref{fig_absorption-armchairs}, indicating two energy scales for optical transitions. In the (12,3) SWNT, which does not have reflection symmetry, the transitions within the doublets are possible even in a clean lattice, causing weak absorption peaks to appear in the low THz range.
Many additional peaks appear at high frequencies for a lattice with vacancies in Fig.~\ref{fig_absorption-armchairs}. Most importantly, the broken symmetries allow also strong low-THz optical transitions to occur. Hence we conclude that our proposed THz detector can be realized with SWNTs of arbitrary chirality.
%
%
%%%%%%%%%%%%
%%%%%%%%%%%%
%
%
%
%
%%%%%%%%%%%%
%%%%%%%%%%%%
%
%
\section{Photon assisted tunneling processes}\label{AppPAT}

We present here a more detailed discussion of the two tunneling processes induced by photon absorption in the Coulomb-blockaded quantum dot with charging energy $E_\text{C}$, for simplicity again omitting the spin. Both the temporary emptying ($N-1$ process) and temporary filling ($N+1$ process) of the dot start from the same initial many-body state, the ground state of the dot with $N$ electrons, where the $N$-th electron occupies the lower, $d$ state of a split doublet. In both processes the first stage is the absorption of a photon and the subsequent excitation of the $N$-th electron to the higher, $u$ state.\\
\begin{figure*}
 \includegraphics[width=\textwidth]{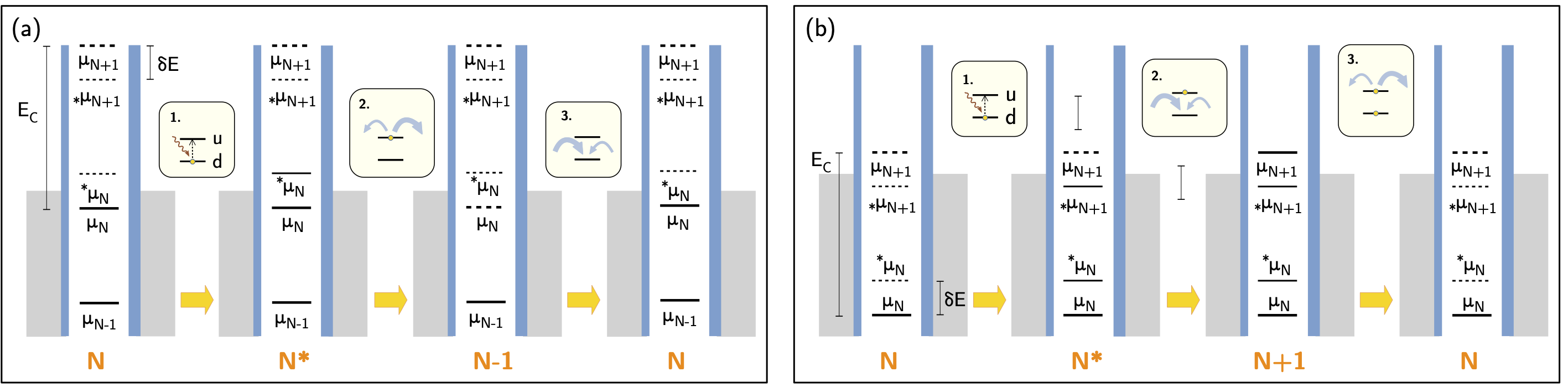}
 \caption{\label{fig:tunneling}
 Energy balance in the three stages of an (a) $N-1$ and (b) $N+1$ tunneling process at zero bias. The tunneling occurs through barriers separating the central QD from the leads, whose continuum of occupied states is marked with grey rectangles, reaching up to the chemical potential of the leads. The chemical potential of the dot at each stage is the highest marked with a solid line; chemical potentials marked with dashed lines correspond to unoccupied many-body states; either they are unreachable by tunneling because of the energy cost (e.g. $\mu_{N+1}$ in the first panel of (b)) or because the QD is not in the required initial state (e.g. $^*\mu_N$, $_*\mu_{N+1}$ in the first panel of (b)).
 }
\end{figure*}
In the $N-1$ process, illustrated in Fig.~\ref{fig:tunneling}(a), in the initial stage the gate voltage is applied in such a way that the chemical potential of the leads is between $\mu_N$ and $^*\mu_N$, and no tunneling can occur. After the photoexcitation the chemical potential of the dot is higher than that of the leads, and the second step, $N^*\rightarrow N-1$ transition becomes possible. Since the chemical potential of the leads is higher than $\mu_N$, in the last step the dot is refilled from the leads, by tunneling into the $d$ state and resetting the device.\\    
In the initial stage of the $N+1$ process shown in Fig.~\ref{fig:tunneling}(b) the $N$-th electron is in the $d$ state, but the chemical potential of the leads lies between $_*\mu_{N+1}$ and $\mu_{N+1}$. The photoexcitation promotes the electron to the state $u$, allowing the dot to be filled from the leads by an $N^*\rightarrow N+1$ transition. Since $\mu_{N+1}$ is again higher than the chemical potentials of the leads, the dot returns to the initial state by a $N+1\rightarrow N$ process, where the $N+1$-st electron tunnels out from the excited state.

\bibliography{cnt}

\begin{thebibliography}{127}
\bibitem [{not()}]{noteLDOS}%
  \BibitemOpen
  \href@noop {} {}\bibinfo {note} {{This is proportional, in first
  approximation, to the local density of states (LDOS), c.f.
  Ref.~[\onlinecite{Tersoff85}]}}\BibitemShut {NoStop}%
\end{thebibliography}

\end{document}